\documentclass[12pt]{article}

\usepackage[margin=1in]{geometry} 

\usepackage{authblk}

\usepackage{latexsym,amsmath,amssymb,graphics,cite,svg}
\usepackage{enumerate}
\usepackage{cancel} 
\usepackage{graphicx}
\usepackage{framed}
\usepackage[colorlinks]{hyperref}
\usepackage{url}
\usepackage{cite}

\usepackage[all]{xy}



\newtheorem{theorem}{Theorem}

\newtheorem{remark}[theorem]{Remark}

\numberwithin{theorem}{section}

\def\be{\begin{equation}}
\def\ee{\end{equation}}
\def\bea{\begin{eqnarray}}
\def\eea{\end{eqnarray}}
\def\ba{\begin{array}}
\def\ea{\end{array}}

\def\bx{{\mathbf {x} }}

\let\<\langle
\let \>\rangle

\newcommand{\rem}[1]{}

\newcommand{\bv}{\boldsymbol{v}}

\newcommand{\pp}[2]{\frac{\partial #1}{\partial #2}}

\newcommand{\abs}[1]{\lvert #1 \rvert} 

\newcommand{\revision}[2]{#2}  
\newcommand{\revisionintext}[2]{#2}  

\begin{document}

\title{The Dictator's Dilemma: The Distortion of Information Flow in Autocratic Regimes and Its Consequences}
\author{Vakhtang Putkaradze}
\affil{  Department of Mathematical and Statistical Sciences, \\ University of Alberta, Edmonton, AB, T6G 2G1, Canada; \\ 
Email: putkarad@ualberta.ca} 

\date{\today}

\maketitle

\abstract{Humans have been arguing about the benefits of dictatorial versus democratic regimes for millennia. Despite drastic differences between the dictatorships in the world, one of the key common features is the \emph{Dictator's Dilemma}  as defined by Wintrobe  \cite{wintrobe2000political}: a dictator will never know the true state of affairs in his country and is perpetually presented distorted information, thus having difficulties in making the right governing decisions. The dictator's dilemma is essential to most autocratic regimes and is one of the key features in the literature on the subject. Yet, no quantitative theory of how the distortion of information develops from the initial state has been developed up to date. 
I present a model of the appearance and evolution of such information distortion, with subsequent degradation of control by the dictator. The model is based on the following fundamental and general premises: a) the dictator governs aiming to follow the desired trajectory of development based only on the information from the advisors; b) the deception from the advisors cannot decrease in time; and c) the deception change depends on the difficulties the country encounters. The model shows effective control in the short term (a few months to a year), followed by instability leading to the country's gradual deterioration of the state over many years. I derive some universal parameters applicable to all dictators and show that advisors' deception increases parallel with the decline of the control. In contrast, the dictator thinks the government is doing a reasonable, but not perfect, job. Finally, I present a match of our model to the historical data of grain production in the Soviet Union in 1928-1940. 
\\
\emph{Keywords}: Dictator's dilemma, Stochastic models, Societal applications, Asymptotic solutions
}
\tableofcontents
\section{Introduction} 
\label{Sec:Introduction} 
Humans have been arguing about the benefits of dictatorial vs. democratic regimes since the dawn of our civilization. Plato \cite{Plato_Republic} has outlined his preference for an Aristocracy, a  government run by an enlightened philosopher king, with incorruptible and enlightened advisors and enforcers. For Plato, Democracy was one of the lowest and the most unjust forms of government, a precursor to Tyranny. Almost 2500 years after Plato, humans have seen the implementation of autocratic governments that both failed and succeeded and democracies that were either successful or gave birth to Tyrannies. Among the common critiques of democracy among the supporters of autocratic government is the short horizon of thought for politicians and the bending of politicians' message to what people generally want to hear, which Plato was already familiar with. Those who favor autocratic regimes praise their abilities to create efficient, quick, long-term solutions for society's problems.    The theme of advertising quick, efficient solutions and governance seems common among the dictators who came to power, especially during relatively recent history \cite{BenGhiat2020strongmen}.

 The literature on understanding totalitarian and autocratic regimes is vast, and we will only attempt a short review here. 
There are certainly many types of dictatorships, from those who take power to enrich themselves and their families to the ideologically driven dictators who perceive themselves as the nation's saviors, righting the historical wrongs and enforcing 'international respect' for their country. Despite the large variety among the dictatorial regimes, 
some common patterns have emerged in the literature that are common to all of them. In particular, in \cite{wintrobe1990tinpot}, Wintrobe coined the term 'the tinpot regime' and developed an economic theory of maximizing the available rent while preserving stability by supporting the allies and security forces. These ideas were further developed in \cite{olson1993dictatorship,mcguire1996economics}, who described the dictators as the 'stationary bandits' who take power to maximize the wealth collection from the population through taxation. The wealth may be in the form of personal riches for the dictator's family if the enrichment is the goal of the dictator, but it could also be understood more broadly as the resources needed to achieve the dictators' goals for the development of the country. For the political economy of dictatorships and balances of repression vs. loyalty and power vs. budget, we refer the reader to \cite{wintrobe2000political,wintrobe2007dictatorship}, see also \cite{wintrobe2013north} for interesting application of these theories to North Korea. In the balance of repression vs. loyalty discussed in these works, Plato's (elusive) Timocracy would correspond to low repression and high loyalty of its citizens and all-too-common Tyranny to high repression and low loyalty. 

Autocratic regimes, stable as they may appear on the surface, need to put effort into their long-term survival using repression and some form of legitimacy obtained from elections. Since we will be concerned with the long-term governance of autocratic regimes, it is worth discussing the survival strategies and the information flow in such societies, in which information flow distortion from the regime to its subject plays an important role. While some dictatorial regimes only allow farce elections, with the incumbent receiving close to 100\% of the vote, some autocratic rulers allow relatively free elections in their country. Egorov and Sonin have considered the advantages of incumbent autocratic candidates in such elections \cite{egorov2014incumbency} and derived the theory of elections where an authoritarian ruler, certain in their popularity, allows relatively free elections \cite{egorov2021elections}. Balancing ruling by a small circle of advisors and getting more widespread support from the population was discussed in \cite{gandhi2007authoritarian}. The paper \cite{wright2012authoritarian} further explains how having parties and legislature in a dictatorial regime increases its legitimacy and prolongs its survival. Further work on the stability of autocratic regimes \cite{gerschewski2013three} identified three pillars of stability: the legitimization of the regimes, repression of the opponents, and co-optation of the allies. The paper \cite{chen2017information} illustrated the role of propaganda and censorship by an authoritarian regime to preserve stability. As important as the top-down propaganda is to the regime's long-term survival, we will focus on the opposite direction of the information flow, namely, the information flowing to the dictator, which also becomes a victim of distorted and broken information flows. 

Having adequate information at their disposal is essential to autocrats' long-term survival, as the absence of accurate information may lead to erroneous or catastrophic decisions. However, the intense use of regime-supporting propaganda distorts the information flows in the country, eroding the truthfulness of information reaching the ruler. Interestingly, resource-poor autocratic regimes tend to allow more freedom to the media as an incentive to improve the quality of government \cite{egorov2009resource}. The situation is exasperated by the fact that autocratic rulers tend to choose their advisors based not just on competence alone but on the balance of competence and loyalty \cite{egorov2011dictators}. Lower-quality advisors are more likely to adorn the truth because of the fear of repercussions. 
History shows numerous cases of this information flow breakdown, the fall of the Romanian dictator Ceausescu being one of the most prominent ones \cite{hardin1997one}. 

The phenomenon of the breakdown of information flow to the dictator has received the name of \emph{The Dictator's Dilemma} using the terminology from the work of Wintrobe \cite{wintrobe2000political}. Since the dictator is assumed to be all-powerful in society, the advisors who help run the country are enticed to flatter his performance and the results of his policies. The dictator may desire objective information but have no way of gathering it. Therefore, the dictator lacks adequate information for decision-making, and the flattery does not guarantee loyalty or truthfulness. Thus, the economic policies of the dictatorship suffer because of the lack of adequate feedback. This information asymmetry, in turn, results in poor economic performance of dictatorial regimes over the long term. The effect of degradation of long-term economic performance, driven by the information asymmetry, has been studied in detail in \cite{papaioannou2015dictator}, who verified the effects of the dictatorship on the economic indicators such as economic growth, inflation, and the quality of institutions. The work \cite{papaioannou2015dictator} called the resulting economic challenges \emph{The Dictator Effect}, attributed, to a large part, to this information asymmetry. In this paper, we quantify the mathematical reasoning behind that information asymmetry and show that the deterioration of information flow is inevitable for quite general underlying assumptions. The dictator's dilemma, stemming from the lack of adequate information, has played an important role in the theory of dictatorial regimes. For example, Francisco \cite{francisco2005dictator} studies that concept to analyze the dictator's choice of the right level of repression without causing the backlash (and defines dictator's dilemma as having to make that choice, which is a bit different than the discussion of pure information flow). Crabtree \cite{crabtree2020cults} analyzes cults of personality established by the dictatorial regimes in light of that concept. Malesky and Schuler \cite{malesky2011single} discuss how dictators obtain information from the elections, which they call the 'Dictator's Electoral Dilemma'. Kerr \cite{kerr2014digital} and more recently Young \cite{yang2023digital} discuss the 'Digital Dictator's Dilemma', namely, using the internet control and AI methods to gain some of the desired objective information about the society not available by other means. 

In spite of the active use of the concept of the dictator's dilemma in the social sciences literature,  quantitative mathematical models for how the phenomenon comes about and evolves in time, up to my knowledge, have not been developed. This paper aims to fill this gap and presents an idealized model addressing the increasing corruption of the information flow and the corresponding decay in control efficiency. While the model does not aim to describe any particular historical figure, it provides insight into the essential instabilities of dictatorial regimes, negating autocrats' perceived governance advantages.

 For simplicity, consider a dictator determined to achieve predefined results for the country they govern. To achieve these results, the dictator plans a path to follow -- for example, reach a particular goal for armament production per year by building new factories, organizing new research, developing expertise in certain fields of science and technology \emph{etc.}
Since all the decisions must come from the dictator, to achieve these goals, especially in numerous fields, the dictator must govern with the help of advisors. In dictatorship regimes, the choice of advisors (more aptly called \emph{viziers}) is a balance between competency and loyalty \cite{egorov2011dictators}. Governing the country depends on the dictator's perception of the country's state; that perception, in turn, depends on the information provided by the advisors. A more sophisticated dictator would try to access different information sources beyond what the closest circle of advisors provides. These efforts, however, can be sabotaged by the circle's closest advisors to prevent the information perceived as undesirable from reaching the dictators, with 'Potemkin's Villages' being the most famous example and a common nickname for these kinds of efforts. 

 The model developed here is based on the concept of Stochastic Differential Equation (SDEs) \cite{oksendal2003stochastic,pavliotis2014stochastic} to determine the interplay of dictator's control to achieve a given goal and the information flow from the advisors. The stochasticity is crucial to the model as it prevents the long-term stability of the dictatorship: a system without noise could actually be controlled in some areas over the long term. Using a mathematical and asymptotic analysis of the model, I show that the distortion of information is not an exception but rather a rule, leading to fundamental consequences for running the country and the well-being of its citizens. Also, surprisingly, the model shows that in spite of the deterioration of the actual performance, the dictator is increasingly unaware of the actual state of the country due to the increasing distortion of the information by the advisors, something that has been observed numerous times in history. The approach developed here can be generalized by extending models to include more complex SDEs, including interactions between multiple goals, input from several advisors, and other more complex features.  

While the author is not aware of the application of the SDEs to this particular aspect of social sciences, the application of stochastic dynamics to social sciences is well-established; see, for example,  the use of the analogs of Ising spin models and their generalizations for describing the opinion dynamics \cite{sznajd2000opinion,sznajd2005sznajd,nyczka2012phase}, see also \cite{jkedrzejewski2019statistical} for the review of the current literature on the subject and the discussion of the relevant Fokker-Planck equation. 
Also, the application of SDEs to the general theory of control is well established \cite{nisio2015stochastic}. The applications of stochastic processes to the deterioration of control mechanisms \cite{nguyen2013deterioration} and system failures \cite{alam1976optimal} are particularly important to our method. We only use the simplest case of such control theory, namely, the pointwise linear feedback control, with more general controllers discussed in Section~\ref{sec:PID}.

\section{Modeling assumptions} 
\label{sec:modeling_assumptions} 
  \paragraph{Goal-setting by the dictator} After coming to power, a dictator would plan several goals that are important in their mind and develop a way how to achieve these goals. This path can be understood as a desired trajectory for various aspects of the country's development.  That trajectory can be the production of certain materials, hardware, or weapons; building houses and roads, measurable improvement in the life of the people \emph{etc.}  This point of the theory is justified by the examples of dictators who came to power in the 20th century \cite{BenGhiat2020strongmen}, presenting what they considered a simple and efficient plan of restoring the country's glory, its economic viability, respect for international stage \emph{etc}. Every development goal underpinning a dictator's reach to power will thus come with a plan for a particular resource, material, or product.

Consider a dictator who would like to govern the country using several measurements $\bx=(x_1,x_2,\ldots,x_n)$ for success. Suppose that the dictator has planned the trajectory $\bar{\bx}(t)$ by pre-planning the path each of the measurements should take. The goal of the dictator and their government is to keep the true trajectory $\bx(t)$ as close to the desired trajectory as possible, \emph{i.e.}, \revision{R1Q2a}{minimize $|\bv|= |\bx(t)-\bar{\bx}(t)|$.} The dictator is naturally facing obstacles to their goals: on each time interval, there is an unexpected forcing throwing the system off equilibrium, such as bad weather, rise in commodity prices, revolt, strikes and other factors impeding the progress towards the goal. The unexpected external forcing at time $t_i$ thus changes $\bv_i=\bv(t_i)$ on each time step as 
\begin{equation} 
\bv_i\rightarrow \bv_i^* = \bv_i +\boldsymbol{\sigma}_i \sqrt{\Delta t_i} Z_i \, , 
\label{v_i_evolution} 
\end{equation}  
where $Z_i$ are random variables that can be taken to be normally distributed, and the factor $\sqrt{\Delta t_i}$ is introduced to make the term on the right-hand-size of \eqref{v_i_evolution} finite when $\Delta t_i\rightarrow 0$. The random part of this equation can be understood, for example, to include an unpredictable part of commodity pricing that includes a stochastic component \cite{schwartz1997stochastic}, or effect of weather on the harvest \cite{hanson1988optimal} (and thus well-being and productivity of population), or other effects which are not predictable. As we shall see, the introduction of random effects is crucial for describing the stability and long-term behavior of the system.  In most of this paper, we shall consider $Z_i$ to have a mean zero since one would imagine that the planning is made using a mean expected values for a particular commodity and not the most optimistic or pessimistic values for that commodity. Then, because of the presence of the amplitude $\boldsymbol{\sigma}_i$, we can take $Z_i$ to be the standard normal, although more general random variables (\emph{i.e.}, normal with non-zero mean, colored noise \emph{etc.}) may be considered. 
\\
In a more realistic scenario, the forcing by a normally distributed variable $Z_i$ in \eqref{v_i_evolution} is inadequate, as there are large amplitude events that may severely affect the system. The influence of such large-scale events, generalizing \eqref{v_i_evolution} and, correspondingly, \eqref{1D_dictator} below, will be presented in Section~\ref{sec:generalized}. In addition to the effect of the noise, that Section also studies the ability of the system to persevere in the face of a finite number of shocks.

 \paragraph{Control} The random noise part will force the system to deviate from the ideal behavior $\bv=\mathbf{0}$. A dictator would introduce a force correction to control that deviation, such as hiring more people to deal with a particular problem or a production slump. The correction is proportional to the deviation $\bv$ and is distributed in time, and is thus given by 
\begin{equation} 
\bv_i^*\rightarrow \bv_{i+1} = \bv_i^* -\mathbb{D}  \bv_i^* \Delta t_i\,
\label{correction} 
\end{equation} 
Here,  $\mathbb{D}$ is a diagonal matrix with positive entries, or, in general, a positive definite matrix satisfying $\bv \cdot \mathbb{D} \bv \geq 0$ for all $\bv$. Taking $\Delta t_i \rightarrow 0$, we arrive at the Stochastic Differential Equation (SDE) \cite{oksendal2003stochastic,pavliotis2014stochastic}
\begin{equation} 
\mbox{d} \bv=- \mathbb{D} \bv \mbox{d} t +  \boldsymbol{\sigma} (t,\bv) \mbox{d} W_t
\label{dictator_ideal} 
\end{equation} 
where $\mbox{d} W_t$ is Ito's noise term \cite{oksendal2003stochastic,pavliotis2014stochastic}. The difference between ODEs and SDEs is in the noise term, making actual solutions depending on the noise realization. Thus, a single solution of a SDE for a given noise realization is not representative; a more consistent approach incorporating all possible realizations of the noise must be sought.  The control term $\mathbb{D} \bv$ provided in this system gives, in the ideal case, the convergence to the ideal state $\bx(t)=\bar{\bx}(t)$. We will consider only the pointwise control term, \emph{i.e.} the term in \eqref{dictator_ideal} proportional to the value of the deviation from desired controllers. In engineering, a PID (Point-Integral-Derivative) controller is also used.  With that approach, the control term in \eqref{dictator_ideal} also contains terms proportional to the time derivative and integral of $\bv=\mathbf{x}-\bar{\bx}$, as we illustrate in \eqref{PID_dictator} in Section~\ref{sec:PID}. We shall not consider these terms in the main body of the paper, as they, in fact, do not make the system stable - but make the analytical description considerably more complicated, as I illustrate in Section~\ref{sec:PID} in the analysis of the stability of PID controller in \eqref{eig_cond}. 

\paragraph{Advisors' influence, information flow and dictator's dilemma} In reality, the information for the control is provided by advisors, who are increasingly afraid to give negative information to the dictator, which is precisely the point of the dictator's dilemma. The error $\mathbf{e}$ generated by the increased reluctance to provide the correct information to the dictator accumulates over time, depending on the deviation of the trajectory $\bx$ from the desired trajectory $\bar{\bx}$. Thus, a realistic  equation for the trajectory, taking into account the increasing breakdown in feedback control due to incorrect information $\mathbf{e}$, is given by 
\begin{equation} 
\begin{aligned}
\mbox{d} \bv &  = \mathbb{D} \left(-\bv + \mathbf{e} \right) \mbox{d} t +   \boldsymbol{\sigma} \mbox{d} W_t
\\
\mbox{d} \mathbf{e} & = - \mathbf{k} \mathbf{g}(  \mathbf{e}, \bv) \mbox{d} t 
\end{aligned} 
\label{dictator_real_reduced} 
\end{equation} 
 Here, $\mathbf{k}=(k_1, \ldots, k_n)$ are a set of positive constants of dimensions having inverse time, and $\mathbf{g} = (g_1, \ldots, g_n)$ are the dimensionless functions describing generation of misinformation by advisors for a particular goal $i$. The notation $\mathbf{k} \mathbf{g}$ simply states that at the $i$-th component, the right-hand side of the second equation of \eqref{dictator_real_reduced} is proportional to $k_i g_i$. 
 As we shall see below, the values $1/k_i$ denote typical deterioration times for the information for a particular goal $i$.  
 
 \paragraph{On non-negativity of the functions $g_i$. } In most of this paper, we will take the functions $g_i$ to be positive in every component $i$. The reason for this assumption is that once an advisor compromises the ability to be truthful, they are unlikely to become more truthful in the future -- an argument which can be formalized as follows. The dictator sees $v_i - e_i$ as the truth, as they are not able to access the true  $v_i$ by the dictator's dilemma. If an advisor rapidly changes the value of $e_i$ to become more truthful for some reason, then the dictator would see that the apparent value $v_i - e_i$ has worsened. A reason or excuse must be found for such worsening of the situation by the advisor, as clearly admitting the truth about the misrepresentation before is out of the question for the advisor.   If an excuse for such a worsening situation can not be found, such a bout of honesty will actually be seen as the incompetence of the advisor by the dictator and will lead to immediate dismissal or worse. Thus, in most of the paper, we assume that the lies of the advisors cannot decrease; in other words, $g_i$ are non-negative. 
 \\ 
 However, exceptions to this case are possible when both $v_i$, $e_i$, and $v_i-e_i$ are of the same order. When $v_i$ is small, the changes in $e_i$ can be hidden in fluctuations in $v_i-e_i$. In such a case, an advisor would have to be very careful and only allow $g_i<0$ close to $\bv=0$. Such a case is considered in Section~\ref{sec:generalized} for a one-dimensional system, and it indeed leads to a more stable system, which is not surprising as more competent advisors lead to better governance. The fundamental question is whether such a competent advisor will actually be selected by the dictator, as the competency of advisors has to be balanced by loyalty \cite{egorov2011dictators}. Thus, while this case of a competent advisor is certainly possible, it is perhaps less likely to occur because of other factors involved in governing a dictatorial regime. 
 \\
 We will consider just one noise term in \eqref{dictator_real_reduced} for simplicity. Note that the advisors do not randomly decide when to lie more or less: their misrepresentation $\mathbf{e}$ increases in all components, depending on the difficulties encountered. 

The noise term $\boldsymbol{\sigma}$ can be either constant or depend on the value $ \bv $, for example, $\sigma_i=\sigma_i^0 + b_i \lvert \bv\rvert$, with $b_i>0$, so the noise and adversity increase with deviation from the desired trajectory. We only consider the constant noise term here for simplicity and only focus on the one-dimensional form of equation \eqref{dictator_real_reduced}, which can be written as: 
\begin{equation} 
\begin{aligned}
\mbox{d} v &  = \alpha  \left(-v + e \right) \mbox{d} t +   \sigma \mbox{d} W_t
\\
\mbox{d} e & =  - k g( e, v ) \mbox{d} t 
\end{aligned} 
\label{1D_dictator} 
\end{equation} 
Here, $g(e,v)$ is a function having the same dimensions as $e$ and $v$ with the properties outlined below,  and $k$ has the dimensions of inverse time. Equation \eqref{1D_dictator} is the simplest case of \eqref{dictator_real_reduced}, stating that at least one goal of the dictator can be treated independently of all others. This assumption is valid when the control provided by a dictator and the information provided by the advisor for that particular goal are independent of other goals. The corruption of information favors one direction of the discrepancy, which we set to be negative. Indeed, producing \emph{e.g.}, more steel than planned ($v>0$) may be considered less of an offense by a dictator than producing less steel than planned ($v<0$), so the advisors would adjust their 'adornments of truth' to the negative direction $e<0$. 

The first term is the control term, forcing $v$ to decrease to $v=e$, the state which dictator perceives as being ideal.  The coefficient $\alpha$ has dimensions of inverse time, with $T_c=1/\alpha$ being the time needed for an order from the dictator to reach the person who will implement that order. The last term illustrates the random difficulties experienced by the country, modeled by the It\^{o}'s noise term, with $W_t$ being a Wiener process -- a standard Brownian motion.  The noise coefficient $\sigma$, taken as a constant, measures the strength of the noise. Variables $v$ and $e$ have the same dimension, and $\sigma$ has the dimensions $v$ or $e$ multiplied by the inverse square root of time. One can, in principle, take $\sigma$ to be variable in time due to external effects, such as increasing difficulties due to the changes in local climate, natural catastrophes, or political pressures from abroad, but we are not going to consider this question in this paper for simplicity. A more generalized version of noise involving shocks to the system is considered in Section~\ref{sec:generalized}.
In the ideal governing procedure, advisors never become corrupt so $k=0$ and $e=0$, and \eqref{1D_dictator} becomes Ornstein-Uhenbeck's (OU) process describing, for example, a noisy relaxation of an overdamped Hookean spring. The expectation value of the solution decreases exponentially in time, and the variance remains finite. Thus, the country will stay on the desired path in that perfect scenario forever.

The time scale $T_e=1/k$ is an advisor's typical 'corruption time'. As we shall see below, the adornment of truth by advisor  $e(t)$ grows approximately as $e^{k t}$. Clearly, the corruption of the advisors should be unnoticed by the dictator, so $T_e \gg T_c$, otherwise the dictator would notice the rapid deterioration of the quality of advisors and quickly change them.  
The important dimensionless characteristic of the system is then the ratio of control $T_c$ and advisor corruption $T_e$ times
\begin{equation} 
K=\frac{T_c}{T_e} = \frac{k}{\alpha} \ll 1 \, . 
\label{K_def} 
\end{equation}  
The condition \eqref{K_def} is a fundamental requirement of the theory, having a profound sociological meaning and also making the mathematical analysis of the equation possible. 
As it turns out, we can estimate the constant $K$ numerically for modern societies from the properties of the model, which we will do below in Section~\ref{sec:initial}. 

The function $g(e,v)$ is a function with the following \emph{Feasibility properties:}  
\begin{enumerate} 
\item With no disturbances to the system, the governing system works perfectly forever. In other words, $(v,e)=(0,0)$ is a critical point of a noiseless equation with $\sigma=0$. Mathematically, this requires $g(0,0)=0$. 
\item  The advisors' misrepresentation of the situation to the dictator can only decrease over time. Once a misrepresentation of facts is presented to the dictator, an advisor cannot take it back as it will diminish their credibility. This requires $g(e,v)>0$ in \eqref{1D_dictator}.
\item When the misrepresentations become large, $\lvert e \rvert \rightarrow \infty$, the rate of increase of $e$ must decrease. Indeed, once the reality is distorted enough, the distortion rate delivered to the dictator should slow down. Thus, as $e \rightarrow \infty$, $g(e,v) \rightarrow 0$. 
\end{enumerate} 
{\bf Main results of this paper.}  In this paper, I show that: 
\begin{enumerate} 
\item There is a fundamental instability in equations \eqref{1D_dictator} for arbitrary function $g(e,v)$ satisfying conditions above, leading to all solutions $(v,e)$ tending to $- \infty$, allowing for arbitrary large deviation from the dictator's goals. 
\item When \revision{R2Q2b}{$v$ decreases to $- \infty$}, the misrepresentation of the situation by advisors $e$ follows $v$, so the dictator's opinion of the government, described by $e-v$, remains much smaller than $v$; analytical approximation for dictator's opinion $e-v$ can be found as well for a general $g(e,v)$; 
\item An analytical approximation to that solution and the variance of fluctuations around it can be found for arbitrary $g(e,v)$; 
\item This analytical approximation of the solution and its statistical properties can be used to match the theory to historical data. 
\end{enumerate} 
 In this paper, we take $g(e,v) \sim \lvert v \rvert$ when $e$ is small. It is a non-essential, although quite general, assumption which does simplify the analysis quite considerably. The results of the initial behavior outlined in Section~\ref{sec:initial} and the computation of escape times  from the stable regime Section~\ref{sec:escape_unstable} will be derived for that particular form of $g(e,v) \sim \lvert v \rvert$ when $e$ is small; whereas all other results, including the nonlinear attraction to $v \rightarrow - \infty$ are derived for arbitrary $g(e,v)$ satisfying the conditions above.
The role  playing by different terms in  \eqref{1D_dictator}  can also be elucidated, revealing their 'physical' meaning. Equation \eqref{1D_dictator} can be viewed as the Ornstein-Uhlenbeck (OU) process with the off-set equilibrium being at $e(t)$ instead of $0$. Since the evolution of $e(t)$ described by \eqref{1D_dictator} is much slower than the fast, noisy dynamics of $v$, one can naively view that system as describing a combined evolution of $e(t)$ and the \emph{expectation value} of $v(t)$. In contrast, the random fluctuations of $v(t)$ about the equilibrium are described locally by the OU process \cite{oksendal2003stochastic}. Thus, when the time scales of control and deterioration gave by \eqref{K_def} are sufficiently different, the function $g(e,v)$ describes the evolution of the expectation values, whereas \eqref{1D_dictator} mostly controls the fluctuations about these expectation values. This is the statement we will quantify in more detail later.  

The Fokker-Planck equation for \eqref{1D_dictator} describing the evolution of the probability distribution $p(t,v,e)$ is written  as \cite{oksendal2003stochastic,pavliotis2014stochastic}: 
\begin{equation} 
\pp{p}{t} = \alpha \pp{}{v} \left( (v-e) p\right)  + k \pp{}{e} \left( g(e,v) p \right) + \frac{1}{2} \frac{\partial^2}{\partial v^2} \sigma^2 p\, . 
\label{FP_eq} 
\end{equation} 
While limited analytical progress can be made directly with \eqref{FP_eq}, especially for a general function $g(e,v)$, we use an approximation of the Fokker-Planck equation to derive the escape time to the unstable regime in Section~\ref{sec:escape_unstable}.

\section{Model analysis and results} 
\label{sec:model_analysis} 
\subsection{Initial stages of a dictatorship} 
\label{sec:initial} 
The dictator comes to power because of the population's unhappiness with the current state of affairs, so the deviation from the desired state is finite: $v(0)=v_0 \neq 0$. In the beginning stages of the dictatorship, the new advisors have not yet compromised the truthfulness of their reports, so it is natural to posit $e(0)=0$ as the initial condition for $e(t)$ in \eqref{1D_dictator}.  For small deviations from the desired trajectory, while the pressure on the advisors to misrepresent the truth is still small, the natural choice of $g$ is to choose $g(e,v) \sim \lvert v \rvert \geq 0$. This is the simplest function that satisfies the conditions for $g(e,v)$ for small $e$. With that choice of the function $g(e,v)$, the rates of growth for the initial instability for $v<0$ can be computed analytically. 
As we discussed, because of general assumptions, there is a difference in corruptability for $v>0$ and $v<0$. In particular,  \eqref{g_func_example} we have $g(e,v) \simeq \lvert v \rvert$ and equations \eqref{1D_dictator} become:
\begin{equation} 
\begin{aligned}
\mbox{d} v &  = \alpha\left(-v + e \right) \mbox{d} t +  \sigma \mbox{d} W_t
\\
\mbox{d} e & =  -k \abs{v} \mbox{d} t 
\end{aligned} 
\label{eq_1D_initial} 
\end{equation} 
When $v$ is not changing significantly during the dynamics, the evolution at the initial stages for $v>0$ and $v<0$ is given by 
\begin{equation} 
\mbox{d} \left( 
\begin{array}{c}
v\\ 
e
\end{array} 
\right) 
= -\mathbb{A} \left( 
\begin{array}{c}
v\\ 
e
\end{array} 
\right) \mbox{d} t + 
\left( 
\begin{array}{c}
\sigma \mbox{d} W\\ 
0
\end{array} 
\right) \, , \quad 
\mathbb{A} = 
\left( 
\begin{array}{cc}
\alpha & -\alpha \\ 
k {\rm sign}(v) & 0
\end{array} 
\right) 
\label{lin_evol} 
\end{equation} 
which is a special case of a two-dimensional Ornstein-Uhlenbeck process. 

For a fixed sign of $v$, the eigenvalues of the matrix $\mathbb{A}$ defined in \eqref{lin_evol} are given by 
\begin{equation} 
\lambda^2 - \lambda \alpha - k {\rm sign}(v) \alpha =0 \, , \quad \Rightarrow \quad 
\lambda_\pm= \frac{1}{2}\left( \alpha \pm \sqrt{ \alpha^2 - 4 k \alpha {\rm sign}(v) }\right) 
\label{char_eq} 
\end{equation} 
Positive eigenvalues of the matrix $\mathbb{A}$ correspond to the stable regime and negative to the instability. 
Thus, the linearized system exhibits instability for $v<0$, and the noiseless system is linearly stable for $v>0$. For the case of $k \ll \alpha$ characterising an efficient dictatorship, the eigenvalues in \eqref{char_eq} are approximated as 
\begin{equation} 
\lambda_+\simeq  \alpha  +  k \rm{sign}(v)\, , \quad 
\lambda_- \simeq   - k \rm{sign}(v)
\label{char_eq_sol_approx} 
\end{equation} 
The case $v<0$ is undesirable for the dictator's governance because of the exponential growth of errors $e$ and deviation from the desired state $v$. Once the solution $v(t)$ reaches negative values, the exponential growth of both $e(t)$ and $v(t)$ occurs: 
$(v(t), e(t)) \sim e^{\lambda_-t} \sim e^{k t}$. Thus, exponential growth is associated with the time scale $1/\lambda_-\simeq 1/k$ where $\lambda_- \simeq k $ is the unstable eigenvalue obtained by choosing the $-$ sign in \eqref{char_eq}.

Thus, there is one stable and one unstable direction in the phase space, the unstable growth rate is approximately equal to $k$.  The time scale $1/k$,  as follows from \eqref{char_eq_sol_approx},  is the typical time for the corruption of the system to occur. Since most democratic societies encourage change of the leadership after at most 8-10 years (after a single or two terms), we can assume that this time frame $96$ months (8 years) is a typical value for the corruption of the system, computed from humanity's historical experience. Proceeding from this estimate,  a solution starting from some general initial conditions will grow by a factor of $N=100$ in $96$ months gives $k \sim 2 \log(10)/96 \simeq 0.048\,  \mbox{ months}^{-1}$. In the rest of the paper,  we take $k \simeq 0.05$ months$^{-1}$ in \eqref{1D_dictator}. While this value is certainly approximate, we believe it is a good order-of-magnitude estimate for the value of parameter $k$. That is the value of $k$ we will use throughout the system.  

 To set the fast time scale, we need to choose the parameter $\alpha$ in  \eqref{1D_dictator}. While there is some uncertainty in the choice of this parameter, strongly depending on the society, we take the typical time scale from the dictator's order to the execution to be about 1-2 months, which is quite an efficient dictatorship. Suppose we take the typical response time to be, for example, two months, then $\alpha=0.5$months$^{-1}$. That is the value used everywhere in the paper; other values of $\alpha$ can be investigated by time rescaling.  

\begin{figure}[ht!]
\centering
\includegraphics[width=1 \textwidth]{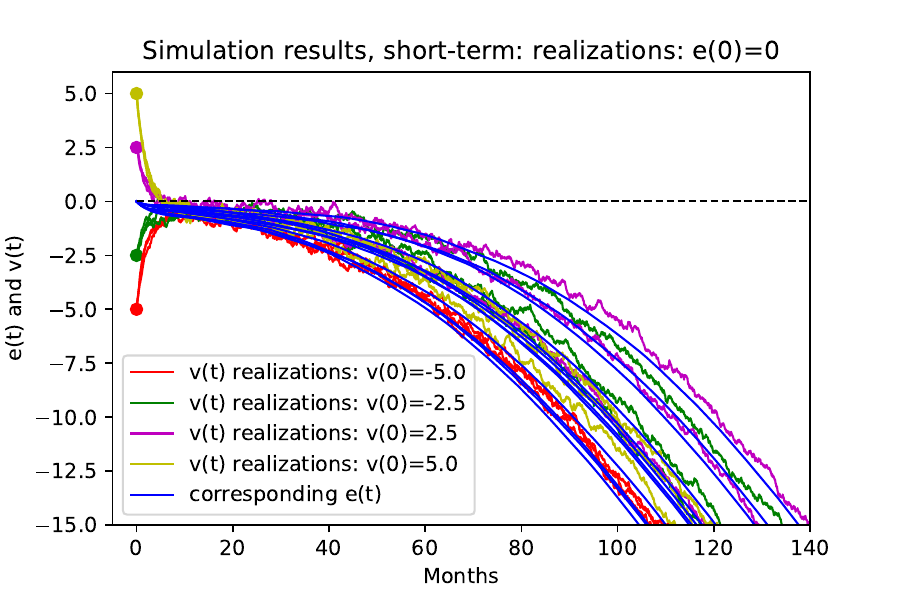}
\caption{
\revisionintext{R1Q2c;R1Q2d}{
Deviation from desired trajectory $v(t)$ (red/green/magenta/yellow lines)  and advisor's misrepresentation $e(t)$ (blue lines) obtained from equations \eqref{1D_dictator} with four initial conditions for $v(0)$: $-5,-2.5,2.5,5$ and $e(0)=0$. The curves of $e(t)$ are drawn with the same color since they remain close to the corresponding $v(t)$ curves. The initial conditions for $v(0)$ are marked by the dots with the corresponding color (red/green/magenta/yellow).  Three realizations of each simulation are shown. The values of parameters $\alpha=0.5\mbox{months}^{-1}$, $\sigma =0.2\mbox{months}^{-1/2}$ and $k=0.05\mbox{months}^{-1}$. This particular simulation is obtained by $g(e,v)$ given by \eqref{g_func_example} with $\mu=1$ and $\gamma=1$.  After the initial decay of the deviation $v(t)$ for the first 10-20 months, the solution for $v(t)$ experiences rapid growth of both  $|e(t)|$ and $|v(t)|$ by a factor typically between 10 and 20, depending on the realization of the solution. Notice that the growth of $|v(t)|$ and $|e(t)|$ occurs simultaneously and $v(t)$ and $e(t)$ remain close while tending to $- \infty$. The simulation time is taken to be equal to 150 months, with the time step $\Delta t=1$ day. The simulation used the \emph{SDEInt} package of Python programming language implementing the Euler-Maruyama algorithm for It\^{o} equations \cite{maruyama1955continuous}. 
}
\label{fig:sim}
 } 
\end{figure}

 To illustrate the system's behavior, we present three realizations of solutions starting with four initial conditions $v_0=(-5,-2.5,2.5,5)$ on Figure~\ref{fig:sim}. The value of the noise coefficient for simulations is chosen to be $\sigma=0.2$months$^{-1/2}$. 
For the calculations presented in this paper, we posit the following form of the function $g(e,v)$: 
\begin{equation} 
g(e,v) =  \frac{\lvert v \rvert}{1+ \mu \lvert e \rvert^\gamma}, \quad \mu>0, \, \, \gamma>0,  
\label{g_func_example} 
\end{equation} 
where $\mu$ and $\gamma$ are some parameters. This function satisfies all the conditions for $g(e,v)$ specified above: $g(e,v)$ is clearly positive, $g(e,v) \sim \lvert v \rvert$, 
and also $g(e,v)$ saturates (\emph{i.e.}, tends to a constant) or goes to $0$ depending on $\gamma$ when $\lvert e \rvert \rightarrow \infty$. In addition, as we show below, the asymptotic solution can be derived explicitly in terms of special functions as a benefit of the functional form \eqref{g_func_example}, although that particular result is not essential for further discussion. The functional form of $g(e,v)$ presented by \eqref{g_func_example} is perhaps the simplest one can find, having the minimum number of parameters and satisfying all the requirements on the function $g(e,v)$.   The results of the paper (except for those of Section~\ref{sec:escape_unstable}) will be valid for any function $g(e,v)$ satisfying the conditions above, with the only difference the asymptotic solution may be expressed in quadratures (implicit solution) rather than explicit solutions that are possible for the choice \eqref{g_func_example}. The results of Section~\ref{sec:escape_unstable} on the escape time are valid for a particular choice of the function, linear in $\lvert v \rvert$; the same is true for Section~\ref{sec:PID}. The details of the escape time from the stable area will of course change when $g(e,v)$ has a more general dependence on $v$ close to $\lvert v \rvert \sim 0$. 
\\
 The meaning of the parameters $\mu$ and $\gamma$ is explained as follows. The physical meaning of $\mu$ can be elucidated by writing $\mu=1/\lvert e_* \rvert^\gamma$ where $e_*$ is some critical value beyond which the advisors have to tone down their misrepresentations, as their information would become less believable should they continue the increase of $e$ at the same rate. The parameter $\gamma$ controls whether the rate of advisors' misrepresentation is increasing ($0<\gamma<1$), decreasing ($\gamma>1$), or tends to a constant ($\gamma=1$) for large $e$. From dimensional considerations, $\gamma=1$ is advantageous; otherwise, $\mu$ would acquire fractional dimensions of $e$ or $v$ for $\gamma \neq 1$. In this paper, we take $\gamma=1$, both because of the dimensional argument above and also because it seems that providing misinformation at a constant rate ("a stream of small lies") would be psychologically satisfactory for both the advisor and the dictator. The parameter $\mu$ is expected to be fitted to the data, whereas $\gamma$ can be either fixed, like we do here, or used as another fitting parameter.

\revision{R2Q2c}{From Figure~\ref{fig:sim}, we observe that after the dictator comes to power, the value of $|v(t)|$ decreases considerably during the first year by roughly an order of magnitude. If the goal was achieved, the particular task would succeed. However, after the initial decay of the undesired deviation $|v(t)|$, that quantity grows again, reaching several times its original value $|v(0)|$ at the final time of simulations of 150 months. More precisely,  $v(t) \rightarrow - \infty$ as $t \rightarrow \infty$ as we will see below. Thus, the initial advantages of the dictatorship, for the chosen values of parameters, are obliterated by the time democratic governments tend to change. The initial decay of $|v(t)|$ is observed due to the particular choice of initial conditions $e(0)=0$ -- no errors in advisors' report to the dictators at the initial time, which is an idealized case. For advisors who start their careers by misrepresenting the state of affairs right away, $e(0)=e_0<0$, and the growth of $|v(t)|$ is observed much sooner and is much more pronounced. The instability persists for more complex controllers like PID (Proportional-Integral-Derivative) \cite{paz2001design} as I show in Section~\ref{sec:PID}. 
}

\subsection{Failure to stay in the stable regime} 
\label{sec:escape_unstable} 
The noiseless system \eqref{1D_dictator} with $\sigma=0$ is stable for all times in the area $v>0$. All solutions with initial conditions starting at $v(0)>0$ with $ e(0)=0$ will forever remain in the $v>0$ domain and will decay exponentially to $0$ when $\sigma=0$. A dictator may be tempted to keep the deviations $v(t)$ in the stable domain $v>0$, guaranteeing the success of their regime. Unfortunately for the dictator, due to noise, it is impossible to keep the system in a stable domain for more than a short amount of time. The escape time is probabilistic and depends on the noise realization for a particular trajectory. The escape time can be estimated from the known expressions for first zero crossings for the Ornstein-Uhlenbeck process \cite{alili2005representations}, and is on the order of months, definitely not enough to ensure a long stable reign. This result can be derived as follows. 

The linearized equation \eqref{lin_evol} becomes a two-dimensional version of the Ornstein-Uhlenbeck process with the formal solution for $\bx =(v,e)^T$
\begin{equation} 
\bx(t) = e^{-A t} \bx(0) + \int_0^t e^{-A(t-s)} \boldsymbol{\sigma} \mbox{d} W_s
\label{OU_sol} 
\end{equation} 
with $\boldsymbol{\sigma}=(\sigma,0)^T$ and $A$ given by \eqref{lin_evol} for $\mbox{sign}(v) =-1$. Because of the noise term, $v(t)$ will cross $v=0$ in finite time. The distribution of the appropriate times can be estimated as follows. 
Suppose $\mathbf{w}_{\pm}$ are the eigenvectors corresponding to the eigenvalues $\lambda_\pm$, and $\mathbb{Q}=(\mathbf{w}_+,\mathbf{w}_-) $ is the matrix composed of these eigenvectors. Since the eigenvalues are distinct and real, the matrix is also real and invertible; let us call the inverse matrix $\mathbb{R}=\mathbb{Q}^T$. By transforming the equation \eqref{lin_evol} to the basis $\mathbf{w}_\pm$, \emph{i.e.}, by taking 
\begin{equation} 
\left( 
\begin{array}{c} 
v 
\\ 
e 
\end{array}
\right) = \mathbf{w}_+ \xi_+ + 
\mathbf{w}_- \xi_- = 
\mathbb{Q} 
\left( 
\begin{array}{c} 
\xi_+
\\ 
\xi_- 
\end{array}
\right) 
\label{w_transform}
\end{equation} 
and noticing that 
\begin{equation} 
\mathbb{R} \mathbb{A} \mathbb{Q} = {\rm Diag} (\lambda_+, \lambda_-) 
\label{QR_identity} 
\end{equation} 
we obtain, in the linear approximation given by \eqref{lin_evol}, the coupled equations
\begin{equation} 
\mbox{d} \xi_\pm = - \lambda _\pm \xi_\pm \mbox{d} t + 
\sigma_\pm \mbox{d} W\, , \quad 
\left( 
\begin{array}{c} 
\sigma_+
\\ 
\sigma_- 
\end{array}
\right) := \mathbb{R} 
\left( 
\begin{array}{c} 
\sigma
\\ 
0 
\end{array}
\right)
\label{lin_evol_transformed} 
\end{equation} 
One can perform a detailed analysis of the probability distribution function for Ornstein-Uhlenbeck's equation as shown in \cite{pavliotis2014stochastic}, Sec.6.4. However, these calculations do not readily lead to a closed-form solution for the first escape time for our problem. We do not need an exact formula for that escape time, only an appropriate expression estimating the typical time to reach the unstable regime, which turns out to happen in months rather than years. We proceed with the estimate as follows. 

 In the first approximation, we notice that $\lambda_+$ is large while $\lambda_-$ is small, so the $\xi_+$ component of the solution will dissipate much faster than $\xi_-$ and converge to a stationary distribution. Since at the initial stage, $e(t)$ remains close to zero, it makes sense to estimate $v(t)$ crossing zero by the first zero crossing of the full solution of equation \eqref{dictator_real_reduced}. That quantity, in turn, is derived as the first zero crossing of the scalar Ornstein-Uhlenbeck process $\xi_-$ defined by \eqref{lin_evol_transformed}. This problem allows for an exact solution of the  corresponding Fokker-Planck equation in the closed form \cite{alili2005representations} :
\begin{equation} 
p(t) = \frac{\lvert x \rvert}{\sqrt{2 \pi}}
\left( \frac{\lambda}{\sinh \lambda t} \right)^{3/2} 
\exp \left( - \frac{\lambda x^2 e^{- \lambda t}}{2 \sinh \lambda t} + \frac{1}{2} \lambda t\right), \quad x= \frac{\xi_-(0)}{\sigma}, 
\label{analytic_sol} 
\end{equation} 
and $\xi_-(0)$ is defined as a second ($-$) component of $\mathbb{R} (v_0,e_0)^T$.

\begin{figure}[ht!]
\centering
\includegraphics[height=0.23 \textheight]{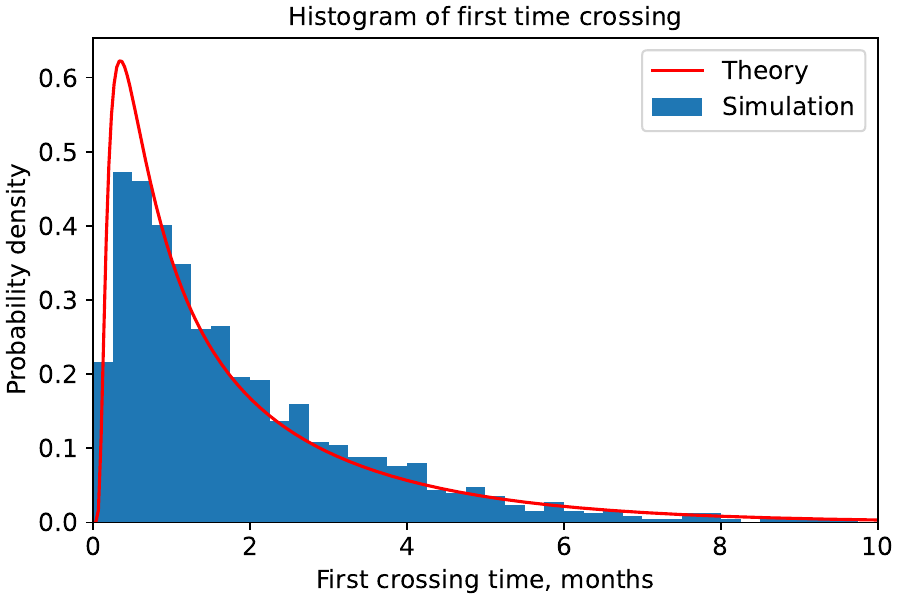}
\includegraphics[height=0.23 \textheight]{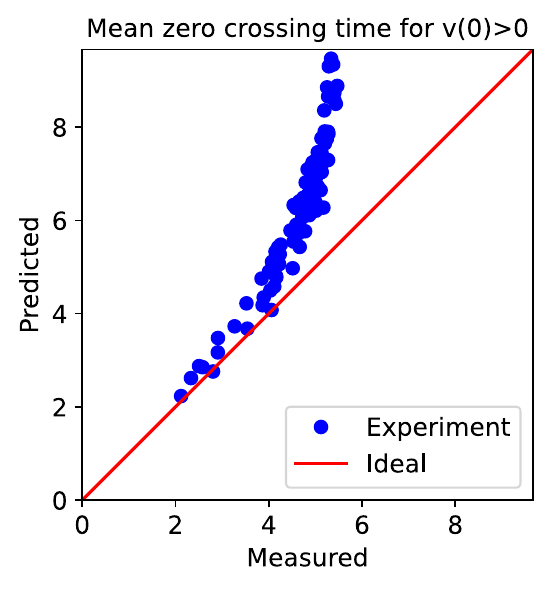}
\caption{Left panel: Distribution of first crossing times for $v(t)$ vs theoretical estimate given by \eqref{analytic_sol}  from \cite{alili2005representations}, starting with initial conditions $v(0)=0.1$, and for the values of parameters $\mu=1$, $\gamma=1$ and $\sigma = 0.2$months$^{-1/2}$. We use the values of $\alpha=0.5$months$^{-1}$ and $k=0.05$months$^{-1}$ as before.  Right panel: Predicted vs. measured average crossing times for 100 examples of the system \eqref{1D_dictator} with $v_0 \in (-0.2, -0.01)$, and $\sigma \in (0.01,0.1)$,  $\gamma \in (0.5,1.5)$, and $\mu \in (0.1,2)$ chosen randomly in the corresponding intervals using a uniform distribution. We perform $1000$ realizations of solutions for each data set to compute the mean time $t$ of the first zero crossings for $v(t)$.  
\label{fig:escape_time} } 
\end{figure}

On the left panel of Figure~\ref{fig:escape_time}, I present a distribution of escape times from the stable $v>0$ into the unstable $v<0$ regime vs. theoretical estimate \eqref{analytic_sol}  adapted from \cite{alili2005representations}. Each one of the 100 data points on this graph is obtained by computing the first crossing time for 1000 realizations with randomly chosen initial conditions and parameters. For the full simulations, the equation \eqref{1D_dictator} with the function $g(e,v)$ given by \eqref{g_func_example} was used with the parameters $\mu=1$, $\gamma=1$, and $\sigma =0.2$months$^{-1/2}$, and, as before, $\alpha =0.5$months$^{-1}$ and $k =0.05$months$^{-1}$. On the right panel of the same Figure, I present the mean value of the escape times versus theoretical estimates of that mean escape time based on the same formula. Note that the escape times from stable to unstable regimes happen in order of several months, which would make any plans by the dictator to keep the system in the stable regime over the long term impossible.  
\begin{remark}[Exact vs approximate results in \eqref{lin_evol_transformed}]
{\rm 
When looking at the transformed equations \eqref{lin_evol_transformed}, one may get an impression that the dynamics of $\xi_+$ and $\xi_-$ can be studied independently. This, however, would only be true if each of the equations would have its own noise term, \emph{i.e.}, $\sigma_{\pm} \mbox{d}W_{\pm}$, and the noise terms $\mbox{d}W_+$ and $\mbox{d} W_-$ are independent. Regarding the Fokker-Planck equation, the diffusion matrix would be diagonal in that case, and the estimate above would be exact. However, in our case, $\mbox{d}W_+ = \mbox{d}W_-$ and the two equations \eqref{lin_evol_transformed} are not independent of each other. This noise dependency is one of the factors contributing to the discrepancy of data from the exact result on the right panel of Figure~\ref{fig:escape_time}. The other factor contributing to the discrepancy is that $e(t)$ deviates from zero sufficiently far for longer times, and $v=0$ is not equivalent to $\xi_-=0$ anymore. Still, the analytical approximation works quite well as an order-of-magnitude estimate for the crossing time into the $v<0$ domain. 
}
\end{remark} 

\begin{remark}[On the use of more general functions $g(e,v)$]
{\rm 
The results of this subsection generalize for an arbitrary function $g(e,v)$ such that $g \sim \lvert v \rvert$ for sufficiently small $v$ and $\lvert e \rvert \rightarrow 0$, such that the estimates based on the Ornstein-Uhlenbeck's approach presented here holds. For a different behavior of function $g(e,v)$ close to $\lvert v \rvert \sim 0$, for example, $g \sim \sqrt{1 + \lvert v \rvert^2}$, or even $g \sim  \lvert v \rvert^2$, the time estimates of this section will not hold. However, for a majority of more general functions $g(e,v)$, there will still be the probability of the escape to the unstable domain $v<0$ in finite time,  although   the actual estimates for the escape time will of course change. We consider $g \sim \lvert v \rvert$ to be the most important example and thus focus our study on that.    }
\end{remark} 
\subsection{Long-term behavior of the system}
\label{sec:long_term}
\paragraph{Asymptotic analysis} Dictators usually come to power with the notion that they bring stability to the country and thus stay for much longer than a typical leader elected in a democratic framework. Thus, it is important to analyze the predictions of our model over much longer time scales than 8-10 years considered above. As it turns out, it is also possible to predict what happens to $v(t)$ and $e(t)$ with high accuracy, which will confirm the results presented in \cite{papaioannou2015dictator} about the detrimental effect of long-term dictatorship on the economic performance due to the information flow breakdown. 

Simulations of long-term solutions show that both $v(t)$ grow indefinitely as $t \rightarrow \infty$. Since large $v(t)$ is the undesirable outcome for the dictator, $e(t)$ needs to be close to $v(t)$ to keep the dictator thinking that the situation in the country is reasonable.

\begin{figure}[ht!]
\centering
\includegraphics[width=0.75 \textwidth]{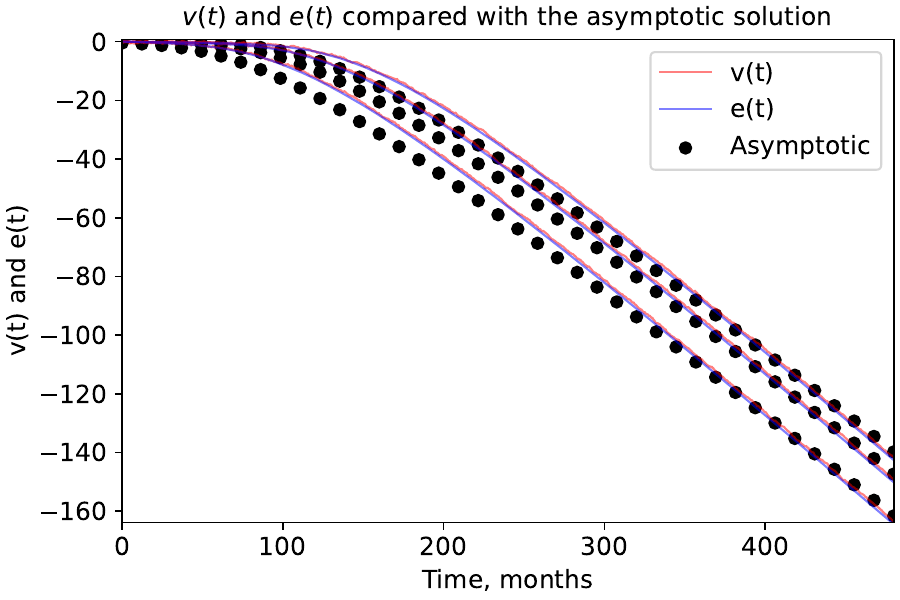}
\caption{Evolution of quantities $e(t)$ and $v(t)$ for several realizations of solution with parameters $\gamma=1$, $\mu=0.1$, $\sigma=0.2$ and initial conditions $v(0)=e(0)=0$, simulated for $40$ years, or $480$ months.  The growth of both $v(t)$ and $e(t)$ are initially exponential, slowing down to a linear growth due to the presence of a denominator in the particular expression for $g(e,v)$ given by \eqref{g_func_example}. The asymptotic solution \eqref{v_sol_explicit} is also shown with dots. The constants $t_s$ in this equation  \eqref{v_sol_explicit} are taken to achieve the best fit to a particular trajectory $v(t)$ and $e(t)$. Three realizations are shown in this Figure for clarity; 100 realizations and asymptotic solution fits were performed in order to study the noisy behavior of the solutions $v(t)$ about the corresponding asymptotic solutions $v_a(t)$. 
\label{fig:long_term_asymptotic} }  
\end{figure}

The approximate asymptotic solution for $v(t)$, $e(t)$ and the difference $u=v-e$ is derived as follows. From \eqref{1D_dictator}, when $v \sim e$, the equation for $v$ reads:
\begin{equation} 
\mbox{d}e = -kg(e,v) \mbox{d} t \quad \Rightarrow \quad \mbox{d} v \simeq g(v,v) \mbox{d} t 
\label{v_eq_approx}
\end{equation} 
leading to the implicit solution 
\begin{equation} 
\int \frac{\mbox{d} v}{g(v,v)} =- k t+C
\label{v_sol_quadrature}
\end{equation}
This  expression is valid for all functions $g(e,v)$ satisfying the conditions on the function $g$. For the particular form of the function \eqref{g_func_example}, further progress can be made. In that case,  \eqref{v_sol_quadrature} becomes: 
\begin{equation} 
\log \abs{v}+  \mu \frac{ v^{\gamma} }{\gamma}
 = k t +C\, . 
\label{v_sol_implicit} 
\end{equation} 
Upon exponentiating and performing some algebra, expression \eqref{v_sol_implicit} gives the expressions for the asymptotic solutions for $v_a$ and $e_a$ as 
\begin{equation} 
v_a = e_a = - W_0 \left( \mu e^{ \gamma k (t-t_s) } \right)^{1/\gamma} 
\label{v_sol_explicit}
\end{equation} 
where $W_0(x)$ denotes the main real root of the Lambert's W-function ($y(x)$ solving  $y e^y=x$ for $x>0$), and $t_s$ is some time shift variable that needs to be computed explicitly for every realization of the solution. 
The comparison of these asymptotic solutions with the full solution of equations \eqref{1D_dictator} is presented in Figure~\ref{fig:long_term_asymptotic}.  Only one realization is shown in this Figure for clarity; a total of 100 realizations of this simulation are performed, together with the best fits to the asymptotic solution.  
The value of the parameter $t_s$ is generally close to the time when the dictatorship began, although we caution the reader against using it directly for that purpose. Indeed, the value of this parameter depends on a particular realization of the noise, as the Figure~\ref{fig:long_term_asymptotic} demonstrates.   

\begin{figure}[ht!]
\centering
\includegraphics[width=0.8 \textwidth]{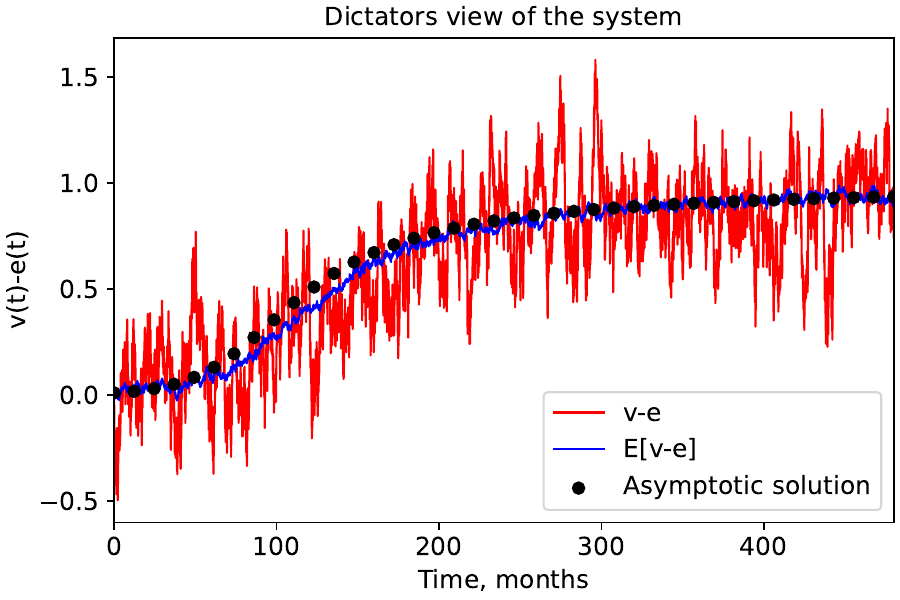}
\caption{The difference $v-e$ is growing much slower compared to either $v(t)$ or $e(t)$, so $v-e$ is several orders of magnitude smaller than the individual values of $v$ or $e$. The expectation value of $v-e$ is computed using $100$ realizations of the simulation of SDEs. The asymptotic solution \eqref{u_approx}   is presented in that Figure with black dots.  That asymptotic solution has no fitting parameters. \label{fig:long_term_u} }
\end{figure}

The difference between $v$ and $e$, which we call $u$,  plays an important role in the system, as that is the value the dictator perceives as the real value of government efficiency. The SDE for $u=v-e$ is computed as follows: 
\begin{equation} 
\begin{aligned} 
\mbox{d} u & = \left( - \alpha u - k g(v-u,v) \right) \mbox{d} t + \sigma \mbox{d} W_t 
\\
&\simeq 
\left( - \alpha u - k g(v,v) +k  \pp{g}{e}(v,v) u + \ldots \right) \mbox{d} t + \sigma \mbox{d} W_t
\end{aligned} 
\label{u_eq}
\end{equation} 
As $v \rightarrow \infty$, it is natural to assume that $\pp{g}{e} \ll 1$ as $t \rightarrow \infty$, so equation \eqref{u_eq} is approximated as 
\begin{equation} 
\begin{aligned}
\mbox{d} u & \simeq \left( - \alpha u - k g_0(t) \right) \mbox{d} t + \sigma \mbox{d} W_t\, \quad g_0(t) = g(v,v) 
\\ 
u & = e^{-\alpha t} u_0 - k \int_0^t e^{-\alpha (t-s)} g_0(s)  \mbox{d} s + 
\int_0^t e^{-\alpha (t-s)}   \mbox{d} W_s
\end{aligned} 
\label{u_eq_approx}
\end{equation} 
where we have defined $g_0(t) = g(v(t), v(t))$. The integral involving $g_0$ in the above expression is approximated for $t \rightarrow \infty$ as 
\begin{equation} 
u (t) \simeq  k \frac{g_0(t)}{\alpha} 
+ k \frac{g_0'(t)}{\alpha^2} + \ldots + 
\int_0^t e^{-\alpha(t-s)} \mbox{d} W_s \, . 
\label{u_approx}
\end{equation}
The non-stochastic part of the solution \eqref{u_approx} is presented on   Figure~\ref{fig:long_term_u} with black dots (filled circles), showing an excellent agreement. Since $g>0$, the asymptotic solution \eqref{u_approx} leads to $u=v-e>0$, so the misrepresentation of truth $e(t)$ by the advisors slightly lags the level of difficulties experienced by the country $v(t)$. 

\begin{figure}[ht!]
\centering
\includegraphics[height=0.24 \textheight]{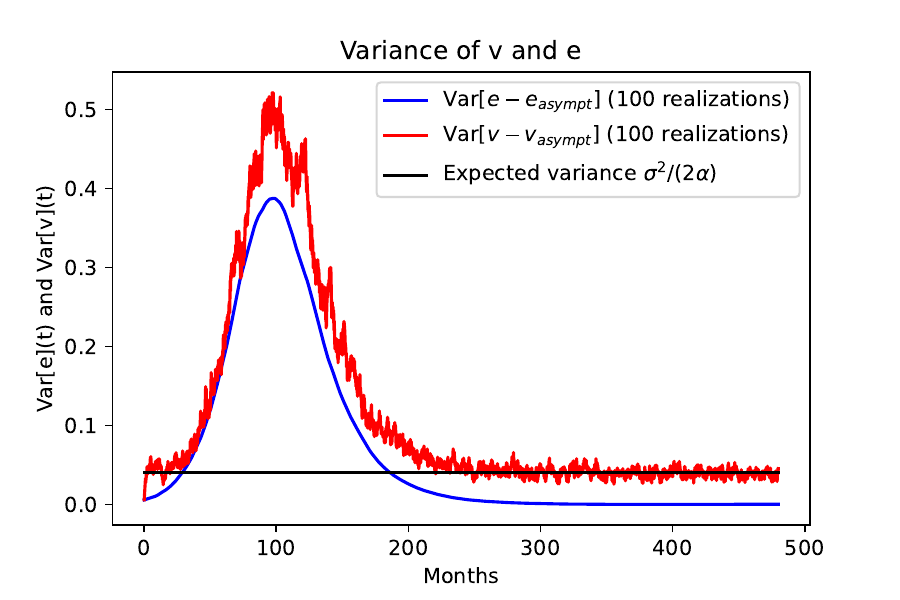}
\includegraphics[height=0.24 \textheight]{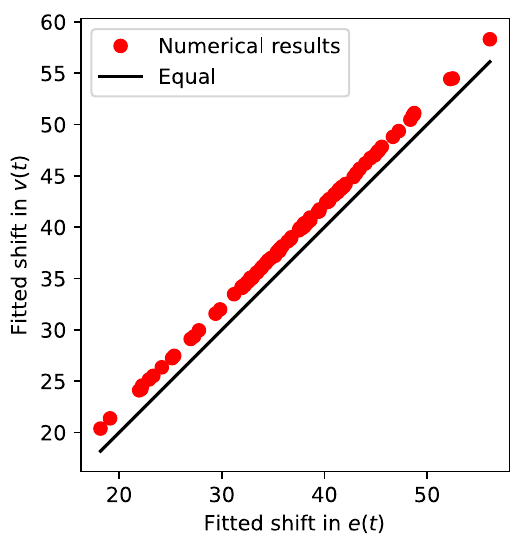}
\caption{ The variance of deviation of $v(t)$ and $e(t)$ from the respective asymptotic solutions, computed over 100 realizations with the same initial conditions as the results on the Figure~\ref{fig:long_term_asymptotic}. For each realization of $(v(t), e(t))$, the corresponding fitting parameters $t_s = (t_v,t_e)$ in \eqref{v_sol_explicit} are determined, the asymptotic solutions $(v_a(t), e_a(t))$ are computed, and the difference between $v(t)-v_a(t)$ and $e(t)-e_a(t)$ is recorded. The variations are then taken from that difference. 
On the right-hand side of that Figure, we present the time shift constants $t_e$ and $t_v$ computed for the system with the same initial conditions. As expected, the fitting constants $t_e$ and $t_v$ are quite close, as is illustrated by their proximity to the $t_v=t_e$ line shown with the solid black line. 
\label{fig:long_term_variance} 
} 
\end{figure}

Next, we investigate how the deviations of the solution $v(t)$ from the asymptotic solution behave with time. Because of the difference in time scales \eqref{K_def}, the evolution of $e(t)$ is relatively slow compared with $v(t)$. From the first equation of \eqref{1D_dictator}, we observe that the evolution of $v(t)$ should be similar to that of an Ohnrstein-Uhlenbeck (OU) process but converging to the asymptotic function $e_a(t)$ (or $v_a(t)$, since they are very close to each other), instead of $0$ or a fixed value like in the standard OU process. Thus, we expect that approximately, the variance of $v(t)$ about the asymptotic solution $v_a(t)$ should coincide with the formula given by the OU process and should be equal to $\sigma^2/(2 \alpha)$. It is indeed the case as Figure~\ref{fig:long_term_variance} demonstrates, showing the variances $v(t)-v_a(t)$ computed over 100 realizations. After an initial increase of the discrepancy due to the mismatch to the asymptotic solution over the short term evident on \eqref{fig:long_term_asymptotic}, the variance of $v$ (red line)
 converges to the expected analytical value  $\sigma^2/(2 \alpha)$ shown with a solid line.   Note the systemic bias between $t_e$ and $t_v$, where $t_v>t_e$, which is due to the fact that for long time, $u=v-e$ is approaching a constant for $\gamma=1$, as the expression \eqref{u_eq_approx} and Figure~\ref{fig:long_term_u} demonstrate. That observed bias will decrease for $\gamma>1$.
 
\begin{figure}[ht!]
\centering
\includegraphics[width=0.9 \textwidth]{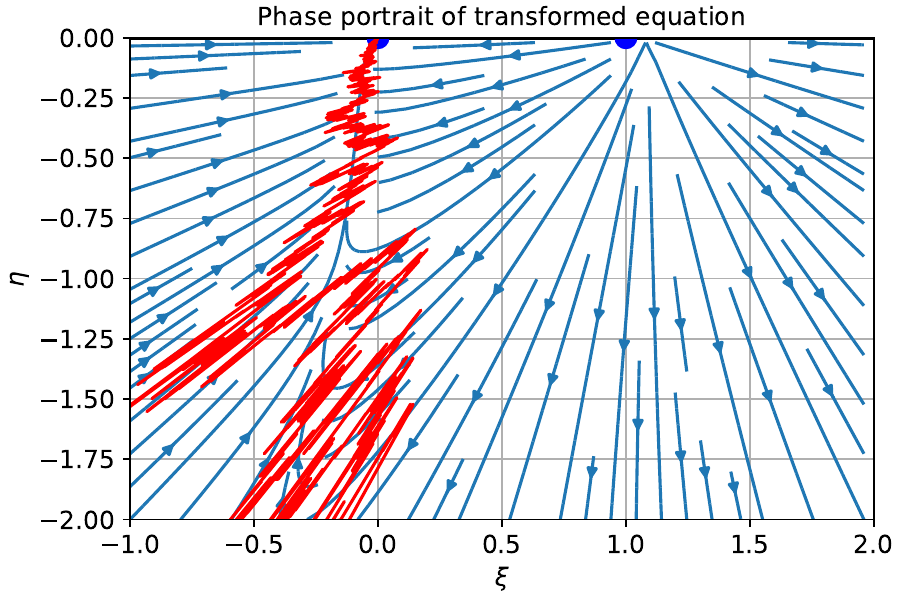}
\caption{Phase portrait of differential equation \eqref{eqs_xi_eta} dropping the terms involving $\mbox{d}W_t$. A particular solution of the original equations $(v(t),e(t))$, transferred into the $(\xi,\eta)$ variables, is shown with a solid red line. That solution converges to $(\xi,\eta)=(0,0)$, corresponding to $v  \rightarrow  +\infty$ and $\abs{v-e}/ v \rightarrow 0$. \label{fig:phase_portrait}  }  
\end{figure}

\paragraph{Phase plane analysis} Let us now show that this behavior of $(v \rightarrow \infty$, $e \rightarrow \infty)$, while $v-e$ remains small, is realized stably on general solutions. In order to do that, we change the equations of motion \eqref{1D_dictator} to the variables 
\begin{equation} 
\xi=\frac{v-e}{v}\, , \quad \eta = \frac{1}{v} \,  , \quad 
G(\xi,\eta) =g\left( e=\frac{1-\xi}{\eta}, v=\frac{1}{\eta} \right) 
\end{equation} 
According to It\^{o}'s theorem \cite{pavliotis2014stochastic}, the equations of motion \eqref{1D_dictator} in the transformed variables are: 
\begin{equation} 
\label{eqs_xi_eta} 
\begin{aligned} 
\mbox{d} \xi &= \left[ - \alpha \xi (1-\xi) + \eta k G - \sigma^2 \eta^2 (1-\xi) \right] \mbox{d}t + (1-\xi) \eta \sigma \mbox{d} W 
\\
\mbox{d} \eta &= \left[ \alpha \xi \eta + \sigma^2 \eta^3 \right] \mbox{d} t - \sigma \eta^2 \mbox{d} W 
\end{aligned} 
\end{equation} 
We can impose the conditions on $G(\xi,\eta)=g(e(\xi,\eta),v(\xi,\eta))$ in \eqref{eqs_xi_eta}, following the conditions on $g(e,v)$ as follows:  
\begin{enumerate} 
\item $G(\xi,\eta)$ is continuous near $(0,0)$ with $G(0,0)=0$ so $(\xi, \eta)=(0,0)$ is the critical point of the deterministic system with $\sigma=0$. 
\item  $G(\xi, \eta)>0$ so the value of the error $e$ is strictly increasing in time. 
\item The term $G(\xi,\eta) \rightarrow 0$ as $(\xi, \eta) \rightarrow (0,0)$ and $G(\xi,\eta)$ is approaching the limit sufficiently fast, so the term $\eta G$ in the first equation of \eqref{eqs_xi_eta} does not contribute to the linear behavior at $(\xi,\eta)=(0,0)$.
\end{enumerate} 
A phase portrait of the equation \eqref{eqs_xi_eta} with no noise term (but still containing terms involving $\sigma$ in the drift terms) is presented in Figure~\ref{fig:phase_portrait} for $\eta<0$. 
All solutions starting sufficiently close to the lower half-plane are converging to the stable critical point $(\xi,\eta)=(0,0)$, with almost a vertical tangent,  which corresponds to $v=1/\eta \rightarrow \infty$ and $(v-e)/v = \xi \rightarrow 0$. The other critical point $(\xi, \eta)=(1,0)$ is unstable. Without the noise ($\sigma=0$), deterministic trajectories are not able to cross the line $\eta=0$. However, with $\sigma>0$, the trajectories starting at $v>0$ become negative relatively quickly, as we discussed above in the short-term evolution of the system, which corresponds to reaching the negative half-plane $\eta<0$ crossing from $\eta = \infty$. A sample trajectory $(v(t),e(t))$, transferred into the $(\xi,\eta)$ variables, is also shown in Figure~\ref{fig:phase_portrait} as a (noisy) solid red line. 

\begin{figure}[ht!]
\centering
\includegraphics[width=1 \textwidth]{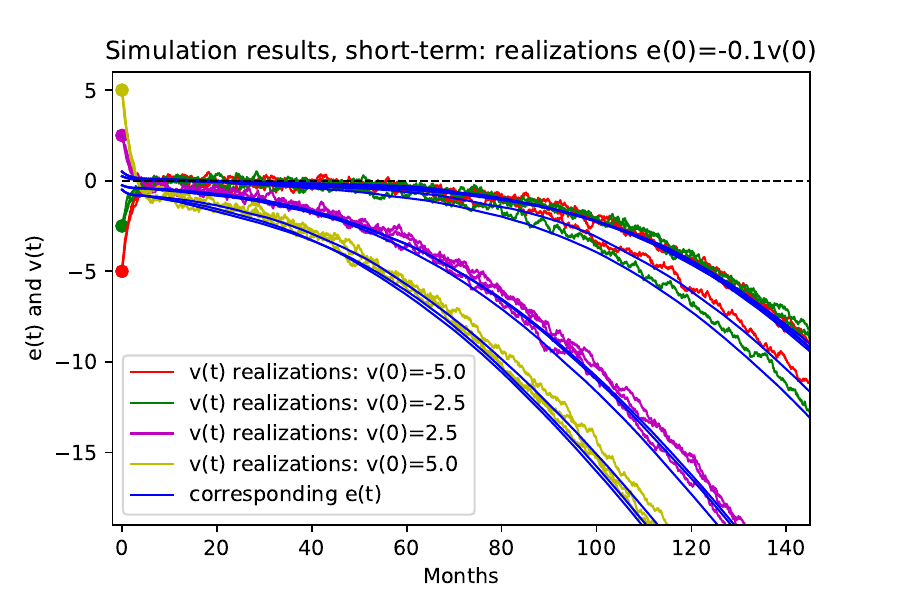}
\caption{ \revisionintext{R2Q2d}{Simulations starting with initial conditions $e(0) = -0.1 v(0)$ corresponding to $\xi(0) =1.1$ in \eqref{eqs_xi_eta}. All other parameters are the same as in Figure~\ref{fig:sim}: $\alpha=0.5\mbox{months}^{-1}$, $\sigma =0.2\mbox{months}^{-1/2}$ and $k=0.05\mbox{months}^{-1}$; with $g(e,v)$ given by \eqref{g_func_example} with $\mu=1$ and $\gamma=1$. 
The red/green/magenta/yellow dots mark, correspondingly, the initial conditions for $v(0)$ for each of the four cases $v(0)=[-5, -2.5, 2.5, 5]$. Three realizations of $v(t)$ for every  one of the four initial conditions are marked with the color corresponding to that initial condition (red/green/magenta/yellow).  The corresponding $e(t)$ is presented with a solid blue line for every simulation, which is always close to the corresponding $e(t)$. There is a slight extension of the stable regime when $e(0)>0$, but eventually, all the trajectories experience unbounded growth $(e,v) \rightarrow - \infty$.  
} 
\label{fig:sim_alt_e} } 
\end{figure}

From the phase portrait presented on Figure~\ref{fig:phase_portrait}, and from the analysis of the transformed equations \eqref{eqs_xi_eta}, one can notice that trajectories starting in the domain $(\xi>1, \eta<0)$ for the deterministic part of the equation \eqref{eqs_xi_eta} (setting $\mbox{d}W=0$) 
do converge to $(\eta = \infty, \xi = \infty)$, which would correspond to the stability of the control. For initial $v(0)>0$, this domain corresponds to the advisor starting to immediately report the additional improvement with $e(0)<0$, and for initial $v(0)<0$, the advisor will choose $e(0)>0$, yielding more pessimistic opinion of the initial state of affairs, but giving themselves some room to improve in the future. 
\\
However, even a presence of such a politically astute advisor does not prevent the instability from happening. On Figure~\ref{fig:sim_alt_e}, we present the results of the simulations for all the parameters being identical to Figure~\ref{fig:sim}, but the initial conditions for $e(0)$ changed to be $e(0) = -0.1 v(0)$, giving $\xi(0) = 1.1$ in \eqref{eqs_xi_eta}. All the solutions start growing eventually, although the stability time may be extended somewhat. Thus, all solutions $(v(t), e(t)$ of the system \eqref{1D_dictator} eventually diverge to $- \infty$, in accordance with the results of Section~\ref{sec:long_term}.

\paragraph{The effect of parameters $\alpha$ and $k$ on the dynamics} 
The value of $k$ will only change the time scales of the long-term instability -- smaller $k$ increases the long-term instability time scale. Increasing $\alpha$ (strengthening the dictatorship) will lower the time scale of short-term initial improvement shown in Figure~\ref{fig:sim}, increases the speed of convergence of $v$ and $e$ to the asymptotic solutions, and diminishes the variance of $v$ around the asymptotic solution. However, the long-term instability of the dictatorship still persists and is controlled by the parameter $k$. Even taking very large $\alpha$ will not save the dictatorship in the long term, although it somewhat stabilizes $v(t)$ and $e(t)$. For example, increasing $\alpha$ by a factor of $10$ lowers the amplitude of growing trajectories on Figure~\ref{fig:sim} by roughly a factor of 2. Of course, a 10-fold increase in the 'harshness' of dictatorship may be cost-prohibitive and unrealistic. On the other hand, the limit $\alpha \rightarrow 0$ would correspond to a 'failed state' (no control) and can hardly be considered a dictatorship. Moreover, such a limit would violate the condition \eqref{K_def}), \emph{i.e.} $k/\alpha \ll 1$, and one would not consider this limit as realistic.  

\section{Instability of a general PID controller} 
\label{sec:PID}
Let us now explore whether a dictator could implement a more general control mechanism to prolong their reign. In this paper, I will consider a standard PID controller to govern the system, with $g(e,v)$ satisfying the same assumptions as the functional form used in Section~\ref{sec:escape_unstable}. The first equation of \eqref{1D_dictator} becomes, instead: 
\begin{equation} 
\begin{aligned} 
\mbox{d} v & = - \alpha_P (v-e) \mbox{d} t - \alpha_D \mbox{d} (v-e) - \alpha_I w \mbox{d} t + \sigma \mbox{d} W_t \, 
\\ 
\mbox{d} w & = (v-e) \mbox{d} t 
\end{aligned} 
\label{PID_dictator}
\end{equation}
where $\alpha_P>0$, $\alpha_I>0$, and $\alpha_D>0$ are the coefficients of the PID controller relating to the point (P), integral (I), and derivative (D) terms, respectively. To observe whether there is a growth of trajectories away from $v=0$, we set $\sigma=0$ in \eqref{PID_dictator} and take \eqref{g_func_example} with small $e$ and write 
\begin{equation} 
\mbox{d} e = - k \abs{v} \mbox{d} t = - k \,  q \, v \, \mbox{d} t\, , \quad q = \mbox{sign} v \, . 
\label{e_small} 
\end{equation}
Consider the case when $v$ has a definite sign, and let us see whether it is possible to have the system \eqref{PID_dictator}-\eqref{e_small} be stable for some values of $\alpha_P$, $\alpha_I$, and $\alpha_D$. Looking for solutions in the form $(v,w,e)=(V,W,E)e^{s t}$ (a method which can be made more rigorous by taking the Laplace transform of the equations \eqref{PID_dictator} and \eqref{e_small}), we arrive to the following linear set of equations for $(V,W,E)$: 
\begin{equation} 
\left( 
\begin{array}{ccc} 
s+\alpha_P + \alpha_D s  & \, \, \, \alpha_I \, \, \, & - \alpha_P - \alpha_D s 
\\ 
- 1 & s & 1
\\ 
k q & 0 & s  
\end{array}
\right) 
\left( 
\begin{array}{c} 
V 
\\ 
W 
\\ 
E
\end{array} 
\right) 
= 
\left( 
\begin{array}{c} 
0
\\ 
0
\\ 
0
\end{array} 
\right) 
\label{VWE_system} 
\end{equation} 
For the system \eqref{VWE_system} to have non-trivial solutions, the determinant of the $3 \times 3$ matrix in \eqref{VWE_system} must vanish, which leads to the condition on $s$:  
\begin{equation} 
s^3 (1+ \alpha_D) + s^2 (\alpha_P + k q \alpha _D) + s (\alpha_I + k q \alpha_P) + k q \alpha_I =0. 
\label{eig_cond} 
\end{equation} 
For stability, all roots of \eqref{eig_cond} have to have the negative real part for $q = \pm 1$. The Routh-Hurwitz stability criterion \cite{routh1877treatise} (see also \cite{bodson2020explaining} for a recent pedagogical explanation) gives a necessary and sufficient condition for stability of $s$ given by \eqref{eig_cond}, based on the sign of coefficients of the polynomial and certain relationships between them. In particular, all the coefficients of polynomial \eqref{eig_cond} must be positive. Since $q=\pm 1$, the stability requirement immediately gives $\alpha_I=0$ from the last term. Consequently, we get $\alpha_P=0$ from $s$-term, and then $\alpha_D=0$ from $s^2$ term, so there is no stable PID controller for \eqref{PID_dictator} with $e(t)$ given locally by \eqref{e_small}. If the advisors could allow themselves to tone back their misrepresentation to the dictator, and \eqref{g_func_example} had no absolute sign, one could indeed make a stable controller for the system. However, it seems unlikely that the advisors would follow that path and reverse their statements of previous successes reported to the dictator.

\section{On non-monotonic increase of deception $e(t)$, shock perturbations and stability}
\label{sec:generalized} 
\begin{figure}[ht!]
\centering
\includegraphics[width=0.8 \textwidth]{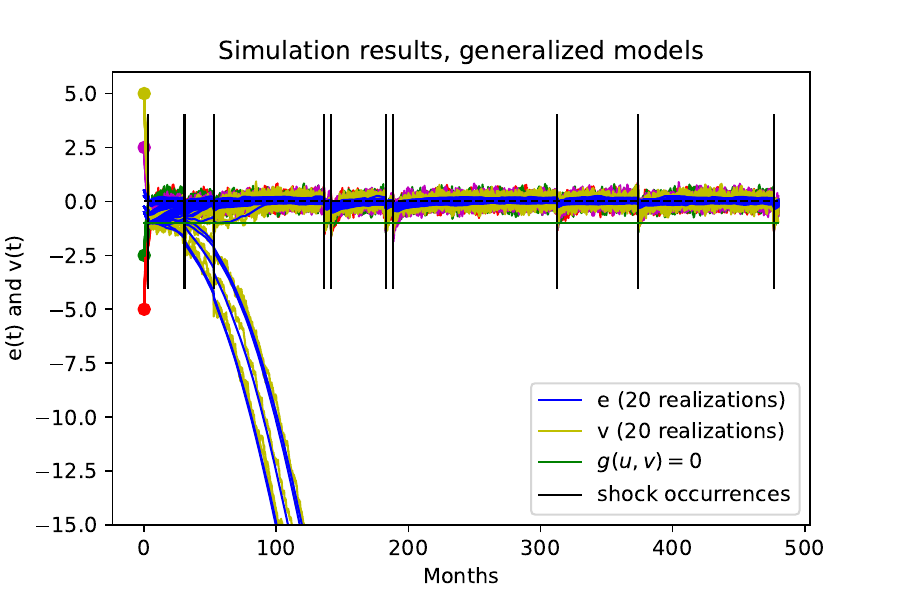}
\caption{\revisionintext{R1Q2d}{Twenty realizations of equations  \eqref{1D_dictator_shocks} with the function $g(e,v)$ given by \eqref{g_func_modified}, with $v_*=-0.5$ and $\zeta =e$ so $g<0$ whenever $-1<v<0$. There are $N_s=10$ shocks experienced by the system during 480 months (40 years) of evolution, on average one every 4 years, each shock having the amplitude $A_v =1$ applied to $v$, and the amplitude $A_e =0.2$ applied to $e$ at the shock. The locations of the shocks are chosen at random during the simulation, taken to be the same for all realizations. The other parameters of the simulations are $\sigma =0.2$, $k=0.05$months$^{-1}$, $\alpha =0.5$months$^{-1}$, as before. Four initial conditions $v_0=[-5; -2.5; 2.5,5]$ are taken, represented by the red/green/yellow dots, with $e_0 = -0.1 v_0$, just as the results presented in Figure~\ref{fig:sim_alt_e}.  The red/green/magenta/yellow lines represent realizations starting rom the corresponding initial conditions. The  solution for $e(t)$ are represented by the blue line; again, $e(t)$ always stays close to the corresponding realization of $v(t)$. }
\label{fig:generalized} }  
\end{figure}

There are several ways to generalize the results here, in particular, allowing for non-positive functions $g(e,v)$ and more general forms of the noise. Let us discuss these generalizations here in more detail, as they may be important for further practical applications of the theory.
\paragraph{Non-monotonic increase of the advisor's deception} One could possibly conjecture that an astute advisor, worried about the system's long-term longevity, may hazard lowering the amount of deception provided to the dictator at certain predefined points. Of course, such lowering of the deception is risky when $v(t)$ is large in absolute value, as it will correspondingly increase the error visible to the dictator, $v-e$, and cause questions about why there was apparent worsening of the situation when there is no apparent reason for it. If a dictator has a good memory, such change in $e(t)$ may also cause questions about the advisor's consistency in providing reports and thus lower the trust in that advisor by the dictator. However, if $v(t)$ is relatively small, non-monotonic fluctuations of the deception may be imperceptible by the dictator. We thus consider a case when a function $g(e,v)$ is allowed to become negative in a certain interval close to $0$. For example, choose some $v_*<0$ and modify \eqref{g_func_example}  as 
\begin{equation} 
g(e,v) =   \frac{ \vert v\rvert }{1+ \mu \lvert e \rvert^\gamma} \varphi(v)\, , \quad \varphi(v) = 1- \zeta e^{- (v-v_*)^2/v_*^2} 
\label{g_func_modified} 
\end{equation} 
If $\zeta>1$, $\varphi (v)<0$ if $v = v_* ( 1\pm \sqrt{\log \zeta})$. In our computations below, we take $\zeta = e$ and $v_*<0$ so $\varphi(v)<0$ for $ 2v_* < v<0$. Several possible values of $\zeta$ and $v_*$ were tested, and the results were similar: the dynamics stabilizes around $v = v_*$ and $e=e_*$ with high probability. While more studies of the system with non-sign definite $g(e,v)$ are needed, it seems plausible that an astute advisor who can manipulate the deception in just the correct way can keep the system stable for a long time and a high probability of success. 
\paragraph{The effect of a shock in the system} The system \eqref{1D_dictator} utilizes the Brownian noise term $\mbox{d}W_t$ to describe the effect of unknown phenomena. Of course, such a system is highly simplified, as the Brownian motion cannot describe the effect of large-amplitude perturbations, such as the imposition of economic sanctions, loss of access to certain materials and technologies from abroad essential for manufacturing, earthquakes, tornadoes, hurricanes, flooding, \emph{etc.}. These unexpected events introduce an instant and finite perturbation in $v$ of the amplitude $A_v$ at a given point $t_e$, the subscript $e$ standing for 'events'. The advisors will need to accommodate that sudden increase of $v$ and 'soften the blow' to the dictator, thus introducing a sudden jump in $e$ of the amplitude $A_e$. One can imagine that $\lvert A_e\rvert  \ll \lvert A_s \rvert $ for every shock occurrence. It is reasonable to assume that $A_v<0$ because it is almost certain that these events will worsen the situation rather than improve it. 
Thus, to account for these extreme events, we introduce the following system to augment \eqref{1D_dictator} 
\begin{equation} 
\begin{aligned}
\mbox{d} v &  = \alpha  \left(-v + e \right) \mbox{d} t +   \sigma \mbox{d} W_t + \sum_{i} A_v^i \delta(t-t_e^i) \mbox{d} t 
\\
\mbox{d} e & =  - k g( e, v ) \mbox{d} t + \sum_{i} A_e^i \delta(t-t_e^i) \mbox{d} t \, , 
\end{aligned} 
\label{1D_dictator_shocks} 
\end{equation} 
where $\delta(t-t_e^i)$  is the Dirac's $\delta$-function positioned at $t=t_e^i$. In the simulations, we take $A_e = 0.2 A_v$ so the deception of the advisors is much smaller than the amplitude of the event itself, and $t_e^i$ are chosen randomly in the whole interval of simulations but kept the same for all realizations. To perform the simulations, we integrate from $t_e^{i-1}$ and $t_e^i$, then reset the initial conditions according to the shock, and repeat for $i = 1, \ldots N_e+1$ with $N_e =10$. Here, $t_e^0$ is taken to be 0, and $t_e^{N_e+1}=T$, and no shock is imposed at the beginning of the simulation or at its endpoint. 
\paragraph{Simulation results} A complete study of the equation \eqref{1D_dictator_shocks} here is beyond the scope of this paper because of the large number of parameters. An example of numerical simulation of this system is presented in Figure~\ref{fig:generalized}, with the values of parameters in the Figure caption. One can see that in some realizations, the system remains stable with both $e(t)$ and $v(t)$ hovering around $v_*$. However, some of the solutions quickly diverge, and no advantage of the 'truthfulness' area $g<0$ is gained. Note that when several shocks are occurring close to each other, the probability of the system going unstable is higher, as is evident from three shocks occurring close to one another at around $t \sim 30-50$ months. Speaking from a general perspective, such a sequence of shocks is actually not improbable, as the corruption and mismanagement in authoritarian regimes often lead to ineffective response to a single unpredictable event, such as an earthquake, and thus follows by \emph{i.e.}, ineffective logistics, local uprisings, and political instabilities. The stability questions, such as the typical percentage of the escaping solutions for a given set of parameters, are highly complex and thus will be undertaken in further studies. For now, it is sufficient to say that an astute advisor who is confident when  operating in the $g(e,v)<0$ regime has a certain probability of success stabilizing the system.   

\revision{R1Q1}{\section{On the information provided by a group of advisors}}
\label{sec:multiple_advisors}
The theory presented here is, in principle, applicable to an arbitrary number of goals (\emph{i.e.}, dimension of the system) and advisors. In this section we will focus on the case of several advisors providing information to the dictator. The dictator will then aggregate the opinions of several advisors and create the control force accordingly. Suppose the advisors provide the inputs $\mathbf{e}=(e_1, \ldots, e_n)$ and the dictator takes into account the advisors' inputs with the weights $\mathbf{w} = (w_1, \ldots, w_n)$, with $\sum_i w_i = 1$. 

Each advisor will provide the information that will be corrupted with the rate $k_i g_i(e_i,v)$. It is natural to assume that the corruption of information is internal, and thus, the increase in deception only depends on that particular advisor's internal judgment. In what follows, in all our computations below, we shall also put for simplicity $g_i(e_i,v) =g(e_i,v)$, with $g(e,v)$ given by \eqref{g_func_example}. Thus, deception is described as a function of the same form for all advisors. The only difference between the advisors' deception functions is contained in the coefficients $k_i$, which we assume to be different for all advisors. 
Then, the following system describes the corruption of information in the case of multi-advisor input: 
\begin{equation} 
\begin{aligned} 
\mbox{d} v & = - \alpha ( v- \overline{e} ) \mbox{d} t + s \mbox{d} W \, , \quad \overline{e} = \mathbf{w} \cdot \mathbf{e} = \sum_i w_i e_i \, , 
\\ 
\mbox{d} e_i & = - k_i g_i(e_i,v) \mbox{d} t +s u_i (\mathbf{e})\mbox{d} t \, . 
\end{aligned} 
\label{system_advisors} 
\end{equation} 
Here, we introduced $u_i(\mathbf{e})$ as the interaction term due to the exchange of opinions, and $s$ is the on/off interaction coefficient:  $s=0$ gives no interaction and $s=1$ gives full interaction. We will set, for simplicity, all weights to be equal, $w_i = 1/n$, although in reality, the dictator may favor some of the advisors vs. others because of trust, better qualifications in the area, or some other reason. We will also consider the case when all $k_i$ are different, as it is extremely unlikely that any two advisors are exact copies of each other. 

Here, we briefly outline how such the interaction term $u_i(\mathbf{e})$ in \eqref{system_advisors} can be modeled and the corresponding conclusions one can draw from these generalizations. In developing these models, we need to remember that autocrats normally employ a very small group of advisors, typically of the order of 10. It is natural to assume that these advisors are in continuous contact with each other and exchange their opinions with each other constantly. For such a small group of advisors, the most appropriate model for the interaction term $u_i(\mathbf{e})$ is probably the Hegselmann-Krause model \cite{hegselmann2002opinion}. In that model, opinions are exchanged (averaged) among the individuals close to each other in opinion values. We will use $u_i(\mathbf{e})$ from this model's reformulation for the continuum case derived in \cite{canuto2008eulerian,canuto2012eulerian}. The 'interaction velocity' term considered in these works due to the exchange of ideas is given by 
\begin{equation} 
\label{correction_e} 
u_i (\mathbf{e}) = \frac{\tau}{I_i}  \sum_j   \left(  e_j -  e_i\right)  f\left( e_i -  e_j \right)\, , 
\end{equation} 
where $\tau$ is the time step of the simulation, the scalar function $f(x)$ is even and decays rapidly enough outside the ball of certain constant $d$, preventing interactions of opinions $e_i$ and $e_j$ that are further than $d$ apart, and $I_i$ is the number of nearest neighbors within the given neighborhood of  $e_i$. We remind the reader that the time step $\tau$ is taken to be $1$ day in all simulations of the paper. An alternative formulation of this model with a non-continuous function $\xi$ was also considered in  \cite{blondel2010continuous}. 
\rem{ In \cite{canuto2008eulerian,canuto2012eulerian} the approximation $I_i = n$ was used, which is, strictly speaking, slightly different from 
The Hegselmann-Krause model requires an average over the neighborhood of $e_i$.    However, it was also noted In \cite{canuto2008eulerian,canuto2012eulerian} that for large $n$, the results are similar to the Hegselmann-Krause model. }
 In what follows, we take 
\begin{equation} 
f (x) = \frac{1}{2} \left( \tanh \frac{x+d}{\epsilon}-  \tanh \frac{x-d}{\epsilon} \right) \, , 
\label{xi_particular_equation} 
\end{equation} 
although that particular choice of function is not important for further consideration. The function $\xi(x)$ defined by \eqref{xi_particular_equation}  is equal to $1$ with the high accuracy when $|x| < d$, then decays exponentially to $0$ with the rate $1/\epsilon$. In the simulation in this paper, using the function \eqref{xi_particular_equation} we can approximate the number of points in the neighborhood of $e_i$ as $I_i \simeq \sum_j f(e_i - e_j)$. 

\begin{figure}[ht!]
\centering
\includegraphics[width=0.465 \textwidth]{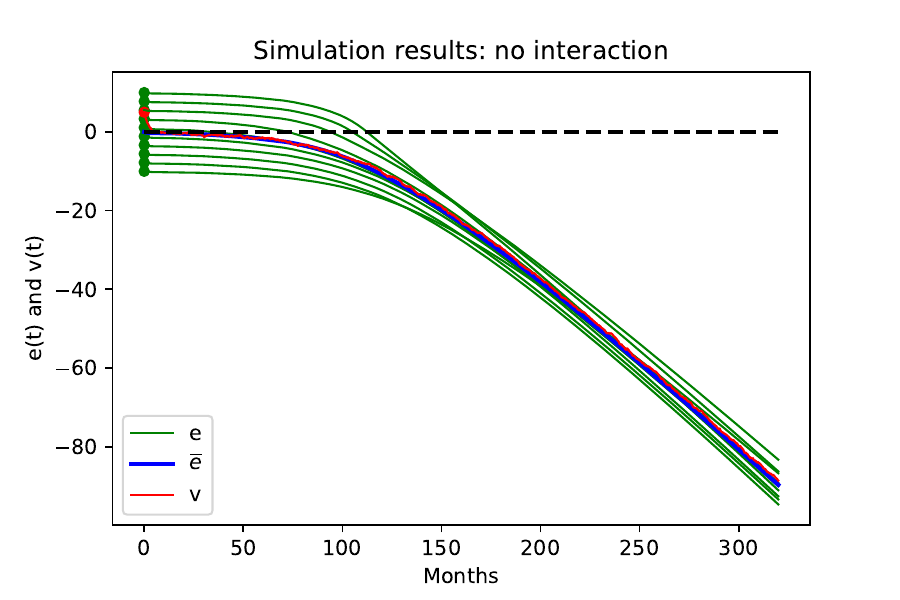}
\includegraphics[width=0.45 \textwidth]{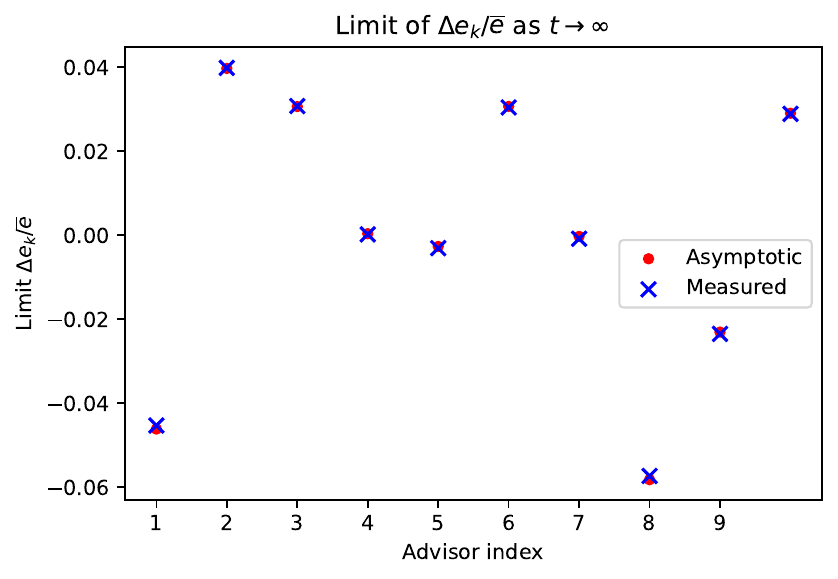}
\caption{
Analysis of solutions of \eqref{system_advisors} with $s=0$ corresponding to an advisory council with no interaction between the advisors. Left panel: a single realization of solutions of equations \eqref{system_advisors}. The green lines are solutions for $e(t)$, the blue line is the average value $\overline{e}$ which is seen by the dictator, and the red line is a realization of $v(t)$. The initial conditions for $e_i(0)$ are equally spaced between $(-10,10)$ indicated with the green dots; the initial condition for $v(0) = 5$ is indicated by the red dot. Both $v(t)$ and $e(t)$ go to $- \infty$ linearly with different slopes, which is due to the functional form of the function $g(e,v)$ chosen for simulations. Note that $\overline{e}(t)$ is always very close to $v(t)$, whereas the opinions of different advisors $e_i$ may be quite far from the mean value $\overline{e}$. Right panel: the results for the asymptotic solution for the slopes of $e_i \sim (A_i t + B_i)t$, \emph{i.e.}, the values of $\beta_i=(A_i - \overline{A})/\overline{A}$. Red dots are the computed asymptotic values of that quantity from \eqref{beta_i_eq}, and blue crosses are the measured values for the relative deviation of the slopes from the mean; the asymptotic values are computed from \eqref{beta_i_eq} for a given set of values $k_i$, $i=1, \ldots n$. The $n=10$ values of $k_i$ are chosen at random from the interval $(0.45,0.55)$  using the uniform distribution. The data presented on the right panel use no fitting parameters. 
\label{fig:no_advisor_interactions} }  
\end{figure}

Before we proceed with the studies of the interaction model \eqref{correction_e}, it is useful to consider the case when the dictator listens to a combination of advisors, but there is no direct interaction between the advisors. As it turns out, that consideration is highly useful for understanding the long-term behavior of the solutions. 
\paragraph{The case of no interactions between the advisors} 
This case corresponds to  taking $s=0$ in \eqref{system_advisors}, \emph{i.e.} dropping the interaction term $u_i(\mathbf{e})$. On Figure~\ref{fig:no_advisor_interactions}, we present only one realization of the solution for clarity \footnote{ All realizations of that  \eqref{system_advisors} look similar, but presenting more than one realization makes the picture hard to read because of an excessive number of lines.}. 
The asymptotic solution of \eqref{system_advisors} with $s=0$ can be derived as follows. We are looking for solutions where $v \simeq \overline{e}$
and, therefore, the asymptotic solutions for $e_i$ are computed as 
\begin{equation} 
\mbox{d} e_i \simeq - k_i g(e_i, \overline{e}) \mbox{d}t  = - k_i g(\Delta e_i + \overline{e}, \overline{e}) \mbox{d}t   \simeq 
- k_i \left( g(\overline{e}, \overline{e}) + \Delta e_i \pp{g}{\overline{e}} (\overline{e}, \overline{e}) + \ldots \right) \mbox{d}t   \, , 
\label{e_asymptotic} 
\end{equation} 
where we have denoted $\quad \Delta e_i = e_i - \overline{e}$ and assumed $|\Delta e_i| \ll |\overline{e}| $ in expansion \eqref{e_asymptotic}. The higher-order term in the parenthesis above is only needed for the functional form of $g(e,v)$ given by \eqref{g_func_example} with $\gamma=1$. We observe from the equations of motion that for $\gamma =1$, all $e_i$ decrease linearly to $- \infty$ and so $\Delta e_i = \beta_i \overline{e}$. For that functional form of the function $g(e,v)$, we have 
\begin{equation} 
\overline{e} ^q \, \frac{\partial^q g}{\partial e^q} (\overline{e},\overline{e}) \sim a_q  g(\overline{e},\overline{e}) , \quad  \mbox{when}  \quad | e| \rightarrow \infty \, , \quad q=1, 2, \ldots 
\label{similarity_g_large_e} 
\end{equation}  
 For all other $\gamma>1$, the terms involving derivatives of $g$ can be dropped in \eqref{e_asymptotic} as they are small compared with $g(e,v)$ as $|e| \rightarrow \infty$. Since we use $\gamma=1$, we will truncate the sum above at the first order term; that accuracy is sufficient for our purposes. A simple computation shows that for \eqref{g_func_example} with $\gamma=1$, the value of the coefficient is $a_1 = -1$ in  \eqref{similarity_g_large_e}. 
Taking the average of \eqref{e_asymptotic} and assuming $\Delta e_i = \beta_i  \overline{e}$ leads to   
\begin{equation} 
\begin{aligned} 
\mbox{d} e_i & =- k_i  g(\overline{e}, \overline{e}) + k_i \beta_i \overline{e} \pp{g}{\overline{e}} (\overline{e}, \overline{e})  \mbox{d}t 
\\
\mbox{d} \overline{e} & = -\overline{k} g(\overline{e}, \overline{e}) + \overline{k \beta} \Delta e_i \pp{g}{\overline{e}} (\overline{e}, \overline{e})  \mbox{d}t \, , \quad \overline{k \beta} := \sum_j k_j \beta_j 
\end{aligned} 
\label{e_asymptotic_average} 
\end{equation} 
Since $e_i = \Delta e_i + \overline{e} = (\beta_i + 1) \overline{e}$, we combine the first equation of \eqref{e_asymptotic_average} and \eqref{similarity_g_large_e}  with $a_1 =-1$ to obtain, as a consistency condition, an algebraic equation for $\beta_i$ for each $i$ 
\begin{equation} 
\frac{k_i -  k_i \beta_i}{1+ \beta_i}  = \overline{k}- \overline{k \beta} \, , \quad i = 1, \ldots n \, . 
\label{beta_i_eq} 
\end{equation}  
The results of the numerical solution of this equation are shown on the right panel of Figure~\ref{fig:no_advisor_interactions}. We measure the fit $e_i \sim (A_i t + B_i)t$ and show the values of relative slopes $(A_i - \overline{A})/\overline{A}$ which in the limit $t \rightarrow \infty$ is the same as $\beta_i = \lim_{t \rightarrow \infty} \Delta e_i/\overline{e}$, and show these values on that panel with blue crosses. We then compute corresponding $\beta_i$ from \eqref{beta_i_eq} and show them on the same panel with red dots. The computed and measured values of the slope show an excellent agreement with no fitting parameters. 

\begin{figure}[ht!]
\centering
\includegraphics[width=0.465 \textwidth]{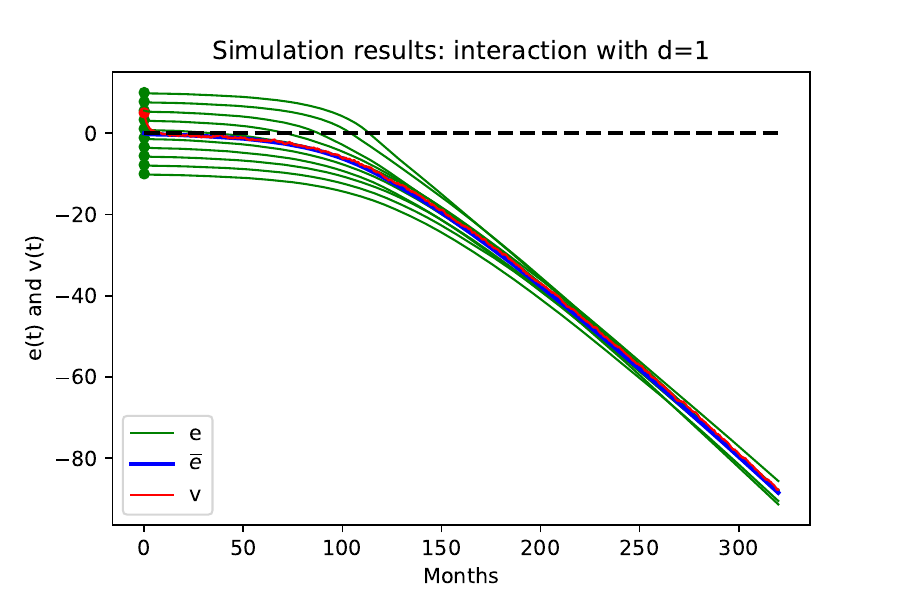}
\includegraphics[width=0.45 \textwidth]{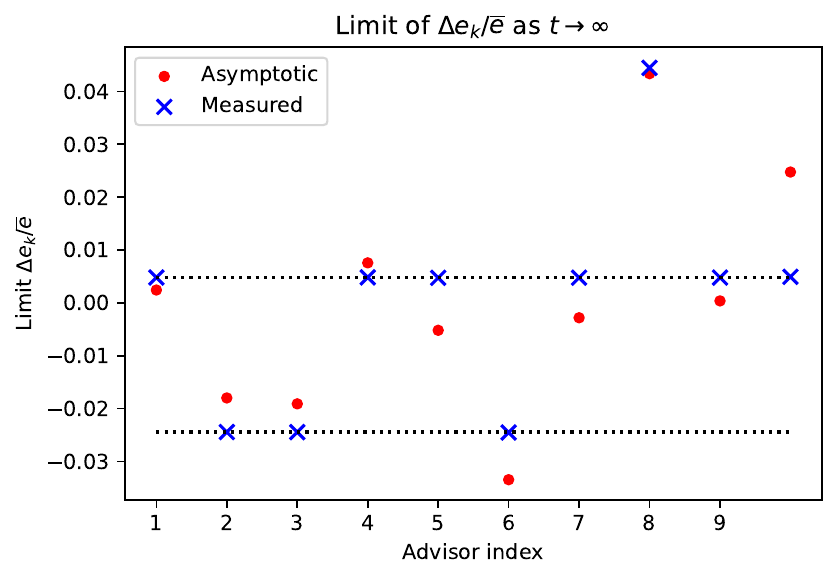}
\caption{
Analysis of solutions of \eqref{correction_e} with $s=1$ corresponding to an advisory board having interactions between the advisors and subsequent clustering of opinions. Left panel: a single realization of solutions $v(t),e_i(t)$, $i=1, \ldots n$ with $n=10$ advisors. The green lines are solutions for $e(t)$, the blue line is the average value $\overline{e}$ which is seen by the dictator, and the red line is a realization of $v(t)$, with the initial condition $v(0)$ marked by the red dot. The initial conditions for $e_i(0)$ and $v(0)$ are the same as in Figure~\ref{fig:no_advisor_interactions}. Both $v(t)$ and $e(t)$ go to $- \infty$ linearly with different slopes, which is due to the functional form of the function $g(e,v)$ chosen for simulations. Again,  $\overline{e}(t)$ is always very close to $v(t)$, while the opinions of different advisors may be quite far from the mean value $\overline{e}$. Right panel: the results for the relative values asymptotic solution for the slopes of $e_i \sim (A_i t + B_i)t$, emph{i.e.}, the values of $\beta_i=(A_i - \overline{A})/\overline{A}$. Similar to Figure~\ref{fig:advisor_interactions}, the red dots on this panel represent the computed asymptotic values for the relative deviation of the slopes from the mean from \eqref{beta_i_eq}, whereas blue crosses are the measured values of the same quantity. The asymptotic values are computed from \eqref{beta_i_eq} for a given set of values $k_i$, $i=1, \ldots n$. There are two clusters of advisors with identical $y$-axis values, formed in this realization indicated with dotted black lines. When clustering occurs, only the non-clustered values are approximated by the asymptotic formula \eqref{beta_i_eq} (advisor number 8 on the right panel in this realization).  The values of $k_i$ are chosen at random from the interval $(0.45,0.55)$  using the uniform distribution, as in Figure~\ref{fig:no_advisor_interactions} (hence a different distribution of red dots compared to the right panel of that Figure). The interaction function $\xi(x)$ is defined by \eqref{xi_particular_equation} with the interaction distance taken to be $d=1$. 
\label{fig:advisor_interactions} }  
\end{figure}

\paragraph{Formation of clusters due to advisor interactions} 
As we observed above, when the interaction without advisors is absent, the difference of opinions of advisors $i$ and $j$, measured as   $e_i-e_j$, diverges without bound, while the average value $\overline{e}$ remains close to $v$. When the interaction between the advisors is present, \emph{i.e.}, the coefficient $s=1$ in \eqref{system_advisors}, there is an initial convergence of opinions due to the interactions. If the terms proportional to $k_i g(e_i,v)$ were not present, the final state for $e_i$ would be a collapse to clusters of several or a single opinion. However, in our case, the initially formed opinion clusters will start diverging linearly from each other due to the interaction with the drift terms $k_i g(e_i,v)$, and the clusters that are distant from each other in the intermediate term will never have a chance to collapse further because of the finite size of interaction $d$. The number of clusters and the placement of the advisors in different clusters is a random process, depending on the realization, and cannot be predicted \emph{a priori}. An example of a realization of a solution with two opinion clusters, shown with dashed black likes on the right panel, is shown in Figure~\ref{fig:advisor_interactions}. On the left panel of this Figure, we show a realization of solutions $v(t), \mathbf{e}(t)$. To further illustrate the formation of clusters, we again measure the fit $e_i \sim (A_i t + B_i)t$ and show the values of relative slopes $\beta_i = (A_i - \overline{A})/\overline{A}$ (blue crosses), versus the computed values of $\beta_i$ from \eqref{beta_i_eq} (red dots). The clustered opinions will have the same value of measured $\beta$. In the case of realization presented here, there are two clustered opinions, indicated by the dashed black lines, and one 'independent' (non-clustered) opinion (advisor 8). As one can see, in the case of the interaction of opinions, the agreement between the asymptotic and measured values of $\beta_i$ is only achieved for the non-clustered opinions. 
Most of the results in opinion dynamics, related to the formation of opinion clusters, were derived for a large number of agents in the system. In our case, the establishment of clusters is due to the aggregating nature of equations \eqref{system_advisors}, which are formally similar to the equations for 'clumpons' without the drift term \cite{holm2005aggregation,holm2006formation}. 

It is also useful to comment on a few other models of interactions between advisors that is possible to implement. If the advisors interact through random encounters,  one could employ the opinion theory by  Deffuant \emph{et al.}  \cite{deffuant2000mixing}. Such a theory would be appropriate if a dictator truly takes advice from a large body, where members have a random chance to interact with each other. An application of that particular model for the dictatorship case considered here is perhaps not realistic but could be useful for other extensions of the theory. If the advisors have to make a discrete choice, \emph{i.e.},    $e_i$ is taking a discrete value of $0$ (no) or $1$ (yes) on a particular subject, then the Sznajd's model \cite{sznajd2000opinion} could also be used. It is reasonable to assume that the advisors do not know the dictator's opinion: if that opinion was known, all advisors would automatically agree with the dictator, and the output from all advisors would be exactly the same. The yes/no answer may also be immune to the pressure to conform to previous answers, as many yes/no questions in governing the country (e.g. whether to build a stadium, a tank factory or a power plant) may be considered independent of each other. The Sznajd model, based on random changes of opinion with certain probability,   is usually analyzed for a large number of agents to demonstrate the convergence to a unique consensus, see \cite{sznajd2005sznajd,nyczka2012phase,jkedrzejewski2019statistical,sznajd2021review,goddard2022noisy}.  For a small number of advisors typically encountered in autocratic societies, the model will experience large values of fluctuations and the convergence to a single opinion is less likely. It is not clear how a dictator will take that consistent lack of universal support for their initiatives: some dictators may find it detrimental, while others may take it as a perceived sign of democracy in their country, legitimizing their reign in their mind. 

\rem{ 
A dictator could, in principle, find advisors who are impervious to the opinions of others and only tell the dictator the truth (in their understanding). However, advisors of this quality are excessively less likely to end up in dictators' courts since, during the selection, the quality of advisors needs to be balanced by their loyalty to the dictator \cite{egorov2011dictators}. 
Even if an incredibly astute advisor could be found, they would be in the absolute minority among the others, their voice overshadowed by the majority of mediocre advisors. The dictator is most likely to listen to that 
majority, although the exceptions to this rule are, of course, possible. In general, modeling interacting advisors who are incentivized to provide an opinion flattering a dictator or aligning with a certain point of view is very interesting and, I hope, will receive further attention in the literature. 
}

\section{Comparison with historical data: grain production in the Soviet Union in 1928-1940} 
\begin{figure}[ht!]
\centering
\includegraphics[width=0.95 \textwidth]{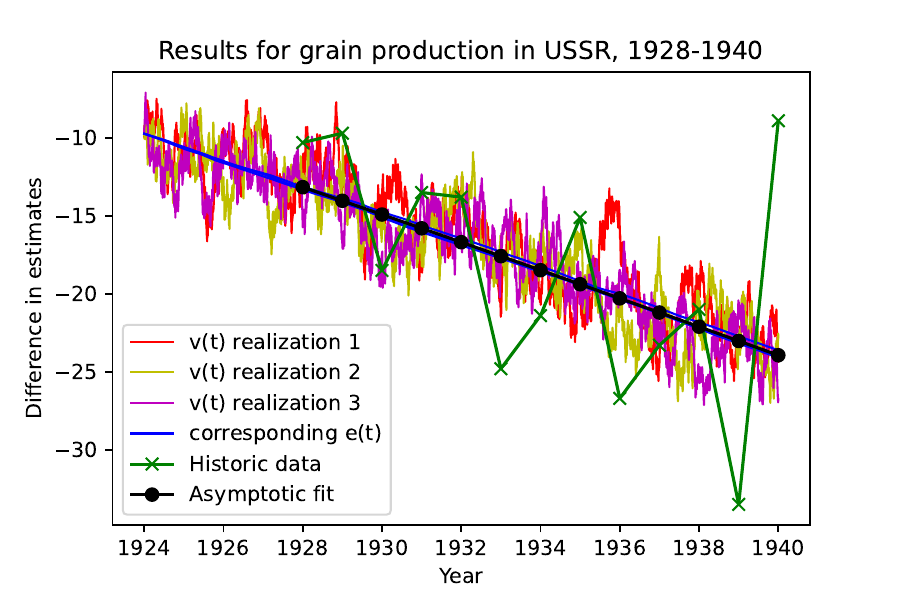}
\caption{Comparision between the difference between the official reports for the Soviet grain production and Western estimates, from \cite{davies1994economic}, and the predictions of our model, shown by green crosses connected by solid green lines. The Least-Square fit of the asymptotic solution \eqref{v_sol_explicit} to the data is presented with black circular markers connected with lines, giving $t_s^* \simeq -164.73$months and $\mu^* \simeq 0.61$MT$^{-1}$.  The asymptotic solution uses, as before,  $\alpha =0.5$ months$^{-1}$,  $k=0.05$months$^{-1}$ and $\gamma=1$.  The noise parameter $\sigma$ is computed to fit the variance of data $V$ to the theoretical value of variance around the asymptotic solution, giving $\sigma= \sqrt{2 \alpha V} \simeq 1.89$MT$\cdot$months$^{-1/2}$.  \revisionintext{R2Q2d}{The red/yellow/magenta line represent three different realizations of the solution, and the  solid blue line shows the trajectory $e(t)$ for every realization. Since $e(t)$ for all realizations are very close, they are shown by the same color. } The initial conditions $(v_0,e_0)$ and the parameter values are the same as in the asymptotic solution. 
\label{fig:experiment} }  
\end{figure}

The challenges of applying quantitative methods such as described here to practical sociological problems is well-known \cite{jkedrzejewski2019statistical}. The application further complicated by the fact that the data provided by autocratic regimes are notoriously unreliable. The data is routinely 'augmented' to the desired values to be used as a propaganda tool to trumpet the regime's successes or conceal the government's failures. Quantifying this augmentation is difficult as only the rosy picture favoring the dictators is presented, whereas the truth tends to be concealed. The 'objective' data for several long-term dictatorships were analyzed in \cite{papaioannou2015dictator} using a linear regression model in log variables for the GDP growth rate, inflation and quality of governing institutions. However, such data cannot be readily used in validating our theory, as it needs both the concealed truth \emph{unknown to the dictator}, and the presented 'augmented' data to quantify the amount of information distortion. Thus, the data satisfying these two combined requirements is quite challenging to obtain. Still, we succeeded in finding historical data that shows a clear discrepancy between the actual and reported data appropriate for applying our theory: the information about grain production in the USSR from 1928-1940. These years have been marked by the Great Famine of the Soviet Union 1932-1933, which affected Ukraine most severely. While there were some environmental challenges related to the harvest around that time, it is believed that the harsh and incompetent local agricultural policy throughout the Soviet Union applied especially brutally to the Ukrainian population for political reasons was to blame, see \cite{tauger2001natural}. The disproportional effect of the famine on the Ukrainian population was recently proven using statistical analysis \cite{markevich2021political}. It is no wonder that the information on grain production during that period was highly politically charged, and the official statistics were changing to reflect the official narrative of the Soviet state. In fact, \cite{tauger2001natural} argues that even the Soviet dictator Joseph Stalin was unaware of the true numbers of grain production during the famine, as illustrated in his own writings from 1932/1933. 
It is likely that the errors in Stalin's own judgment on the matter persisted during the whole time between 1928-1940, passed on by inaccurate reporting that aimed to present the communist agricultural policies as successful rather than report the truth and face the consequences of the wrath of the dictator. In \cite{davies1994economic}, Table 19, one finds official figures for Soviet grain production from that period, the revised figures from the late Soviet Union, and Western estimates about the grain production, which are inevitably lower. The official grain production is then taken to be the desired state at the points of time $t_i$, and the difference between the official and Western estimates is taken as $v(t_i)$, expressed in Millions of Tons (MT). That difference is presented in Figure~\ref{fig:experiment} together with the theory with the parameters fit in the following way.   
\begin{enumerate} 
\item A least-square fit for the asymptotic solution \eqref{v_sol_explicit} with the original values used in this paper $\alpha =0.5$ months$^{-1}$, and $k=0.05$months$^{-1}$, and $\gamma=1$, as discussed in Section~\ref{sec:initial}, is performed to determine the values of the parameter $\mu$ in \eqref{g_func_example} and the shift value $t_s$ in the asymptotic solution \eqref{v_sol_explicit}. The results of the fit are: $t_s^* \simeq -164.73$months and $\mu^* \simeq 0.61$MT$^{-1}$. 
\item Using the computed values, we determine the variance of the data $V$ from the asymptotic solution $v_a(t; t_s^*, \mu^*)$. Using the expression  $V=\sigma^2/(2 \alpha)$, we determine the variance to be $\sigma^* \simeq 1.81$MT$\cdot$months$^{-1/2}$. 
\item Using the asymptotic solution, we compute the initial conditions for $v(t_0)$ and $e(t_0)$ as $v_a(t_0; t_s^*, \mu^*)$, where $t_0=1924$ is the year of Stalin's assumption of power in the Soviet Union. The obtained numerical values are $v(t_0) = e(t_0) \simeq -9.73$MT. 
\item All the parameters and the initial conditions of the equation \eqref{1D_dictator} are now determined, and several realizations of the solution of SDE \eqref{1D_dictator} are performed. Three of these realizations are shown in Figure~\ref{fig:experiment}. 
\end{enumerate} 
Note that all the parameter values are fitted using deterministic formulas, utilizing the theory behind the asymptotic solution \eqref{v_sol_explicit} and stochastic dynamics around that solution. Should one have chosen a different function $g(e,v)$, an implicit solution using quadratures \eqref{v_sol_quadrature} would be used for fitting; however, the general fitting procedure remains exactly the same, as long as $g(e,v)$ satisfies the conditions outlined in Section~\ref{sec:modeling_assumptions}. 
The value of the parameter $\mu^*\simeq 0.61$MT$^{-1}$, corresponds to the advisors beginning to show caution in representing the errors to the dictator at about $e_* \simeq 1/\mu^* \simeq 1.63$MT, which corresponds to about 2.5\% of the total harvest or 16\% of the actual discrepancy between the lie and the truth. That number sounds reasonable -- an advisor would certainly have to tread more carefully when misrepresentation to the dictator reaches that value.  The value of $\sigma\simeq 1.89$MT$\cdot$months$^{-1/2}$ also seems reasonable - these are uncertainties of the harvest corresponding to a few percent of the desirable or actual value. As the theory suggests, this value of $\sigma$ creates exactly the variance of the solution of SDE corresponding to the data; therefore, the variance for all three realizations of SDEs shown (and,  in fact,  all noise realizations one would make) with fitted parameters align exactly with the variance of the data.  

\section{Conclusions} 

  The main contribution of this paper lies in developing a simple yet robust model addressing several aspects of dictatorship control and information flow. The paper's results show that the control in dictatorial regimes exhibits an essential instability caused by the fear of advisors to disappoint the dictator. Another contribution of the paper lies in the fact that stochastic forcing is essential to show the existence of such instability; without the stochastic forcing, there are large stable areas in the phase space where one may think the dictators may contain the system for an arbitrarily long time. Due to the fundamental principles of SDEs, such control can only be achieved for a short amount of time, and long-term control in the stable domain is not possible. Finally, it is interesting that the derived system demonstrates that the dictator inevitably becomes unaware of the true state of the country and believes that the situation is well under control by the government, whereas the opposite is true.

A model of information flow in a dictatorship, presented here, may feel too simplistic to a cautious reader. Nevertheless,  this simple theory explains the initial success of dictatorship and also shows that a dictator can't achieve long-term goals because of the way the evolution of advisors occurs over the years, trying to avoid, as much as possible, the necessity to bring the bad news to the dictator. The theory uses the pointwise control, \emph{i.e.}, the term $- \alpha(v-e)$ in \eqref{1D_dictator}.  The instability persists if a dictator implements a more general controller involving the rate of change and integral (PID controller), as shown in Section~\ref{sec:PID}. The inevitable rise of $v$ should lead to the unhappiness of the people and eventually could lead to a coup d'\'{e}tat or democratic elections. It would be interesting to see whether it is possible to incorporate the criteria for regime stability and resource allocation \cite{guriev2015modern,guriev2020theory}. Some of the dictators still receiving accurate data may even institute a freer information exchange in society \cite{egorov2009resource}, which will, in our model, lead to the decrease of $v$ and an improvement of the economic situation. 
 Several extensions of the model considered here are possible. One could study the more general multi-dimensional problem \eqref{dictator_real_reduced} instead of the one-dimensional problem \eqref{1D_dictator}, which would be of interest, especially if the same advisors address several dimensions in the system. A second direction of studies will be to continue generalizing the functions $g(e,v)$ and the forcing combining the stochastic forcing and shocks, as presented in Section~\ref{sec:generalized}, and derive more detailed analysis of stability depending on the parameters of $g(e,v)$ and the noise/shock forcing.  The third avenue of study would be to have more sophisticated expressions for the parameters. For example,  if the dictator wants to make their reign more repressive, they may try increasing the control coefficient $\alpha$ in \eqref{1D_dictator} as a function of $v-e$; our preliminary investigations show that the system exhibits finite-time singularities, which can presumably be interpreted as the breakdown of the regime. 
Any generalization of the theory, no matter how simple or complex, must obey several fundamental principles: 
\begin{enumerate} 
\item The control must be authoritarian, \emph{i.e.} all control is wielded by the dictator 
\item Since the dictator only has the information provided by the advisors, the control has to be \emph{subjective}: variables can only depend on $(v-e)$ or their multi-dimensional generalizations (\emph{i.e.}, not $v$ or $e$ separately) 
\item All the external influences, such as noise term, must depend on variables that are \emph{objective}, \emph{i.e.} they can depend on $v$ but not on $e$. 
\end{enumerate} 
Finally, it would also be beneficial to consider a variety of advisors interacting with the dictator regarding a single-goal trajectory. In that case, equation \eqref{1D_dictator} would have $e(t)$ in the first equation resulting from a complex dynamics of opinions between several advisors. The dictator may then either average the opinions or listen predominantly to some advisors he considers most truthful, with only a small weight allocated to others.

If a dictator would like to govern for a long time and minimize $v(t)$ over their reign, according to this model, they would need to implement changes that would sound Utopian when applied to a realistic dictatorial regime.  First, a dictator should let the competent advisors who know the real state of affairs govern, in which case the controller term in \eqref{1D_dictator} would be dependent on $v$, not $v-e$. Second, the dictator should create an atmosphere where the advisors are allowed to take back their statements without repercussions, so the function $g$ is in \eqref{1D_dictator} is allowed to change sign, and no $\abs{v}$ in \eqref{g_func_example} is needed. Third, a dictator should choose a representation of opinions and listen to them all instead of a narrowly defined group of advisors without allowing the opinions to converge. If these goals are impossible to achieve, a dictator would need to remove themselves from power in a much shorter time than typically associated with autocratic regimes, consistent with the time frame of term limits in democratic countries.  An autocratic regime following these recommendations would be closer to an ideal Plato's Aristocracy than an actual autocratic regime that existed anywhere in the world. I am thus inclined to conclude that all realistic autocratic regimes are inherently unstable and eventually lead to the degradation of society. 
\section*{Acknowledgements} 
I am indebted to the patience and help of my colleagues and friends who have contributed to discussions and provided suggestions, critique, and encouragement for this article:  
Anthony Bloch, George Constantinescu, Pascal Finette,  Fran\c{c}ois Gay-Balmaz, Darryl D. Holm, Andrea Klaiber-Langen, Brent Lindquist, Svetlozar Rachev,   Feodor Snagovsky,  Magdalena Toda, Dimitri Volchenkov, Peter Vorobieff,  Niall Whelan,  Dmitry Zenkov. I am also grateful to Prof. Jan Luiten van Zanden for the discussion of the preprint of this manuscript and attracting my attention to the paper \cite{papaioannou2015dictator}. This work was partially supported by the National Science and Engineering Research Council of Canada Discovery Grant NSERC RGPIN-2023-03590. 

\subsection*{Author contribution statement} 
The author performed all the work related to this article, including literature search and analysis, derivation of the model, numerical simulations, and theoretical analysis. 
\subsection*{Declaration of interest} 
The author declares no competing interests.

\subsection*{Data Availability} Data for Figure~\ref{fig:experiment} were taken from \cite{davies1994economic}, Table 19. The 'Historical data' points in the Figure are computed as the difference between the Soviet estimate and the low Western estimate from that table. The programs used in this paper are available in \url{https://github.com/vputkaradze/Dictator-equation-programs}.

\subsection*{Ethical approval } 
This article does not contain any studies with human participants performed by any of the authors.

\subsection*{Declaration of AI use}
AI-assisted technologies were not used in creating this article.

\rem{ 
\section*{Ethical statements} 
\subsection*{Ethical approval } 
This article does not contain any studies with human participants performed by any of the authors.
\subsection*{Informed Consent  } 
This article does not contain any studies with human participants performed by any of the authors.
} 
\bibliographystyle{unsrt}
\bibliography{references}  

\end{document}